\documentclass[aps,prd,superscriptaddress,showpacs,twocolumn,nofootinbib]{revtex4}
\usepackage{graphicx}
\usepackage{dcolumn}
\usepackage{bm}
\usepackage{natbib}
\usepackage{amsmath}
\usepackage{amssymb}

\newcommand{\be}{\begin{equation}}
\newcommand{\ee}{\end{equation}}
\newcommand{\bea}{\begin{eqnarray}}
\newcommand{\eea}{\end{eqnarray}}

\newcommand{\WMAP}{{\slshape WMAP~}}
\newcommand{\PLANCK}{{\slshape Planck~}}
\newcommand{\WMAPc}{{\slshape WMAP}}
\newcommand{\PLANCKc}{{\slshape Planck}}

\begin{document}

\title{Energy Injection And Absorption In The Cosmic Dark Ages}

\author{Tracy R. Slatyer}
\email{tslatyer@ias.edu}
\affiliation{School of Natural Sciences, Institute for Advanced Study, Princeton, NJ 08540, USA}


\begin{abstract} Dark matter annihilation or de-excitation, decay of metastable species, or other new physics may inject energetic electrons and photons into the photon-baryon fluid during and after recombination. As such particles cool, they partition their energy into a large number of efficiently ionizing electrons and photons, which in turn modify the ionization history. Recent work has provided a simple method for constraining arbitrary energy deposition histories using the cosmic microwave background (CMB); in this note, we present results describing the energy deposition histories for photons and electrons as a function of initial energy and injection redshift. With these results, the CMB bounds on any process injecting some arbitrary spectrum of electrons, positrons and/or photons with arbitrary redshift dependence can be immediately computed.
\end{abstract}


\maketitle

\section{Introduction}

Precise measurements of the cosmic microwave background (CMB) anisotropies (particularly by \WMAP  \cite{Komatsu:2010fb} and the upcoming \PLANCK Surveyor \cite{Planck:2006uk}) have the potential to probe the detailed ionization history of the universe between recombination and reionization. In particular, dark matter (DM) annihilation or similar new physics can inject high-energy electrons and photons into the photon-baryon fluid, which give rise to additional ionization and heating as they cool, and consequently leave an imprint in the CMB \cite{Padmanabhan:2005es, Galli:2009zc, Slatyer:2009yq}.

High-energy electrons in the early universe dominantly lose their energy by upscattering CMB photons, and the cooling time for photons -- either upscattered or injected directly -- can easily be comparable to a Hubble time \cite{1989ApJ...344..551Z, Chen:2003gz, Slatyer:2009yq}. Consequently, the effect on the ionization history and gas temperature does not precisely track the energy injected; particles injected at a particular redshift may lead to an increase in the residual ionization fraction at a considerably later time.

In this note we employ code developed to map an energy injection history into an energy deposition history, and described in detail in \cite{Slatyer:2009yq}. We compute the deposition histories for a large grid of photon and electron injection energies and redshifts, and make the results publicly available online, at \texttt{http://nebel.rc.fas.harvard.edu/epsilon}; interpolation on this grid can be used to compute the energy deposition history for an arbitrary spectrum and variation of the energy injection with redshift.

We describe our numerical results in Sec. \ref{sec:numerics}, and discuss their interpretation in Sec. \ref{sec:discussion}: as a sample application, we demonstrate the effect of including DM halo formation at low redshifts on the energy-deposition history. In Sec. \ref{sec:review}, we briefly review the principal-component method for estimating CMB constraints developed in \cite{Finkbeiner:2011dx}; in Sec. \ref{sec:constraints}, we outline some applications, showing as examples how our results can be applied to set new and updated constraints on scenarios with late-decaying species, and models of DM annihilation where the annihilation switches on with some characteristic long timescale. In Sec. \ref{sec:conclusions} we present our conclusions; Appendix \ref{app:download} describes the online supplemental material.

Throughout this work, we assume the cosmological parameters from \cite{Larson:2010gs}: explicitly, $\omega_b = 2.258 \times 10^{-2}$, $\omega_c = 0.1109$, $A_s$($k$=0.002 Mpc$^{-1}$) $= 2.43 \times 10^{-9}$, $n_s = 0.963$, $\tau = 0.088$, $H_0 = 71.0$ km/s/Mpc.

\section{Description of Numerical Results}
\label{sec:numerics}

The code developed for \cite{Slatyer:2009yq} takes as an input some injection of photons and electrons, with a specified redshift and energy dependence. Backreaction on the CMB photons and gas is \emph{not} included, as large modifications to the ionization history or CMB spectrum are ruled out by observational constraints, and consequently the problem is (to a good approximation) linear. We thus populate individual energy bins with electrons/positrons or photons at a specific redshift and track the spectral evolution with redshift. Our 40 energy bins are log-spaced between 1 keV and 10 TeV, in photon energy and electron kinetic energy. We employ 65 log-spaced redshift bins spanning the range from redshift 10 to 3000.

At each timestep, the photon spectrum is updated with the results of the various scattering and pair production processes described in \cite{Slatyer:2009yq}, and redshifted, and photons at sufficiently low energies are tagged as ``deposited'' and removed. The threshold for this ``deposition'' occurs when the photon would on average photoionize an atom once per timestep, and as in \cite{Slatyer:2009yq}, we choose a timestep of $d \ln (1+ z) = 10^{-3}$ (it was confirmed in that work that the results were converged at such a timestep). Below this energy deposition threshold, a prescription for the copious secondary electrons from ionization and excitation is needed to translate the energy deposition history to an ionization history. Such prescriptions have been developed in \cite{1985ApJ...298..268S, Valdes:2009cq,MNR:MNR20624}; for the present problem we restrict ourselves to studying the redshift-dependent partition between deposited energy and free-streaming high-energy photons.

In order to improve the speed of the code, we have simplified the treatment of electron cooling relative to  \cite{Slatyer:2009yq}. For high-energy electrons ($\gtrsim 1$ keV), inverse Compton scattering on CMB photons is by far the dominant cooling mechanism. As the electron energy falls, the electron deposits more of its energy to ionization, excitation and heating, but also the energy of photons upscattered by the electron falls (as the square of the electron energy), and these low-energy photons are promptly absorbed by the gas. Thus, in regimes where the initial electron does \emph{not} promptly deposit $100\%$ of its energy, and the partition is non-trivial, the vastly dominant energy-loss mechanism is inverse Compton scattering; for the purposes of this note, we can therefore ignore the other energy loss processes. We have verified that making this approximation changes our results at less than the percent level.

Since we are primarily interested in charge-neutral sources of energy injection, we will usually consider injecting $e^+ e^-$ pairs rather than electrons alone. This is irrelevant for the energy-loss mechanisms important for high-energy particles, but when the positrons have cooled far enough, they annihilate with ambient electrons, producing gamma rays which can in turn escape rather than depositing their energy. We also present results for the injection of electrons only; these can be obtained directly from the results for $e^+ e^-$ pairs and photons. Since the electron cooling time is many orders of magnitude faster than the photon cooling time, it is a good approximation that the positron annihilates at the same redshift it was injected, so an injected positron is equivalent to an injected electron + additional photons at 511 keV and below.

Our results are expressed as a three-dimensional grid for each of the particle types, describing the fraction of the original particle's energy deposited in any given timestep, for a particular redshift-of-injection, and a particular initial energy. We will denote the elements of this grid as $T_{e^+e^-, e^-, \gamma}^{ijk} = T_{e^+e^-, e^-, \gamma}(z_\mathrm{inj}^i,z_\mathrm{dep}^j, E^k) d \ln(1+z_\mathrm{dep})$, where $z_\mathrm{inj}$ is the redshift of injection, $z_\mathrm{dep}$ is the redshift of deposition, and $E$ is the initial energy of the particle. 

The files are available online in \texttt{.fits} and \texttt{.hdf} format, at \texttt{http://nebel.rc.fas.harvard.edu/epsilon} (see Appendix \ref{app:download} for further details); we also supply a Mathematica notebook demonstrating how to read the \texttt{.hdf} files and reproduce the calculations in this note.

\section{Discussion}
\label{sec:discussion}

\subsection{Total energy deposition}

\begin{figure*}
\includegraphics[width=.327\textwidth]{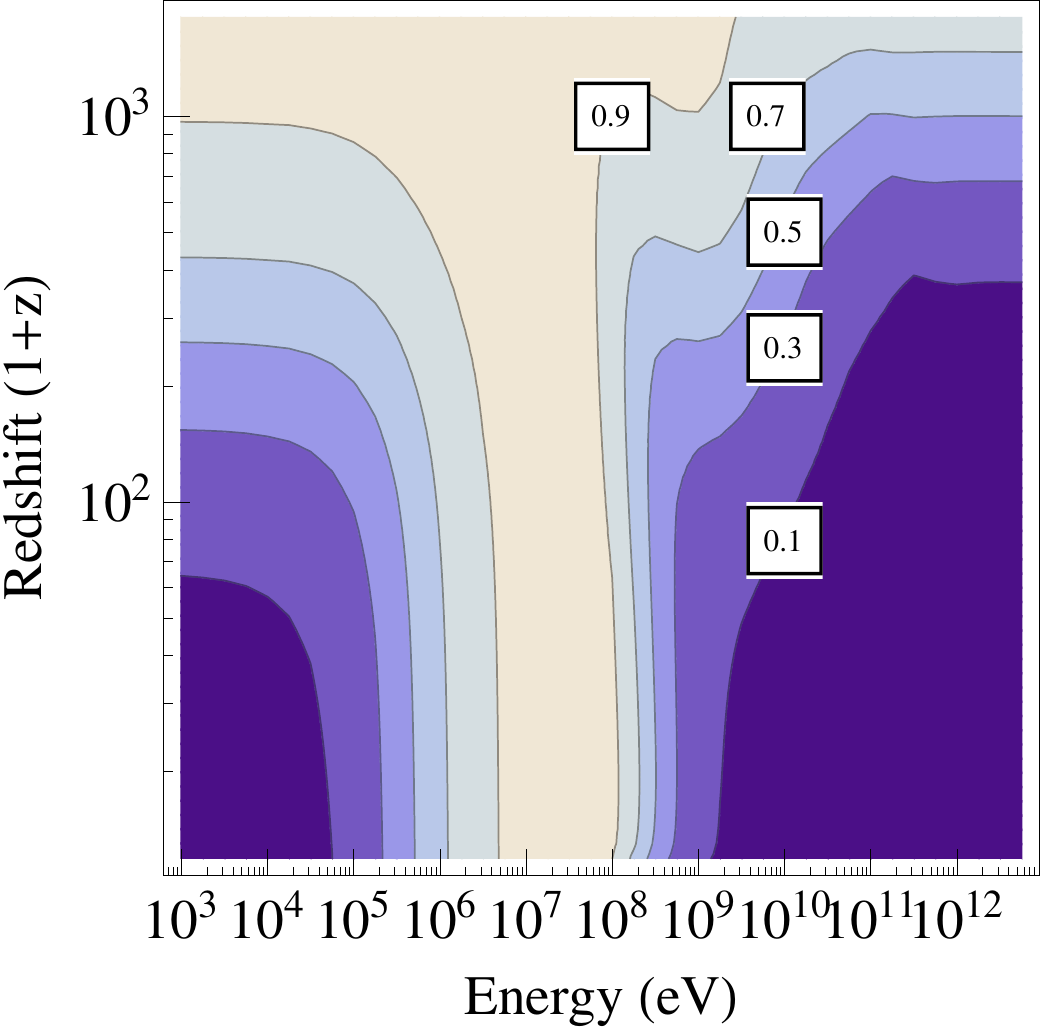}
\includegraphics[width=.327\textwidth]{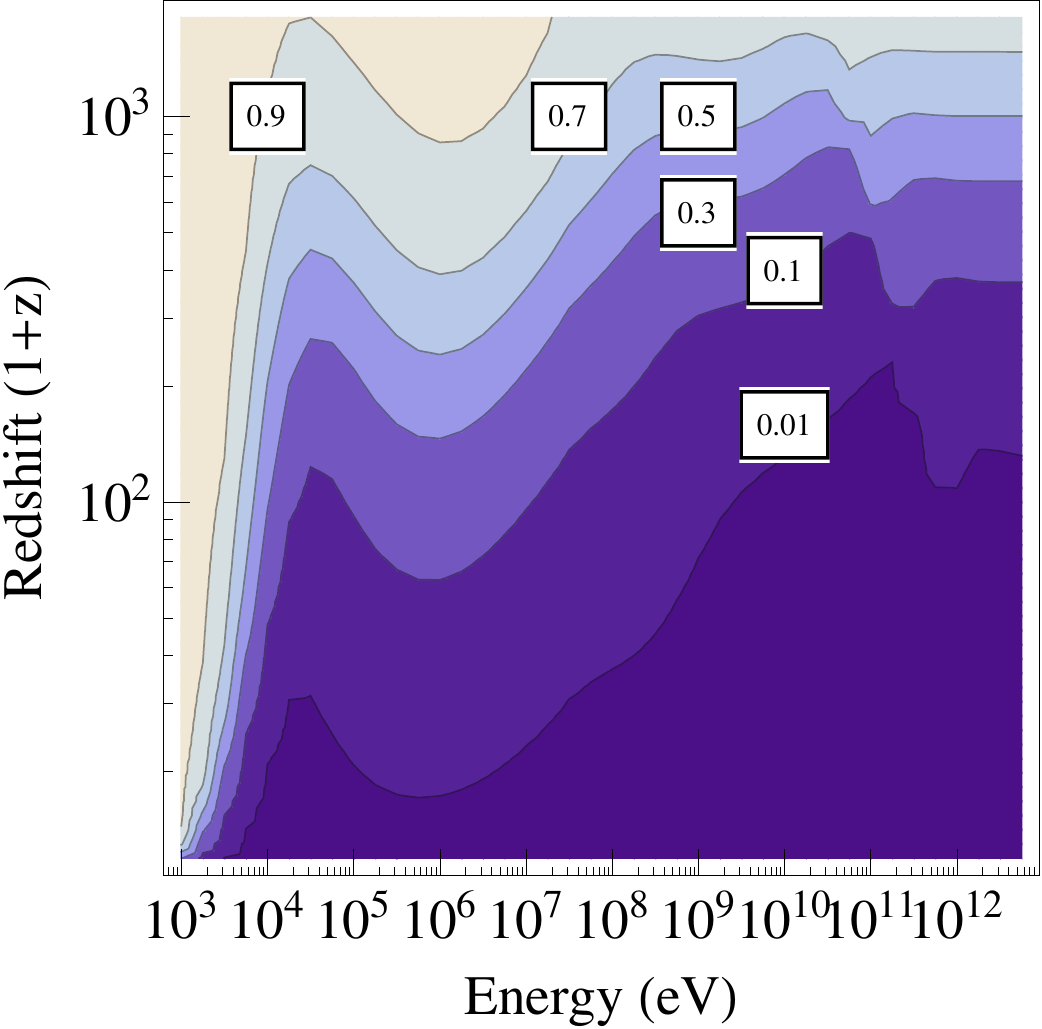}
\includegraphics[width=.335\textwidth]{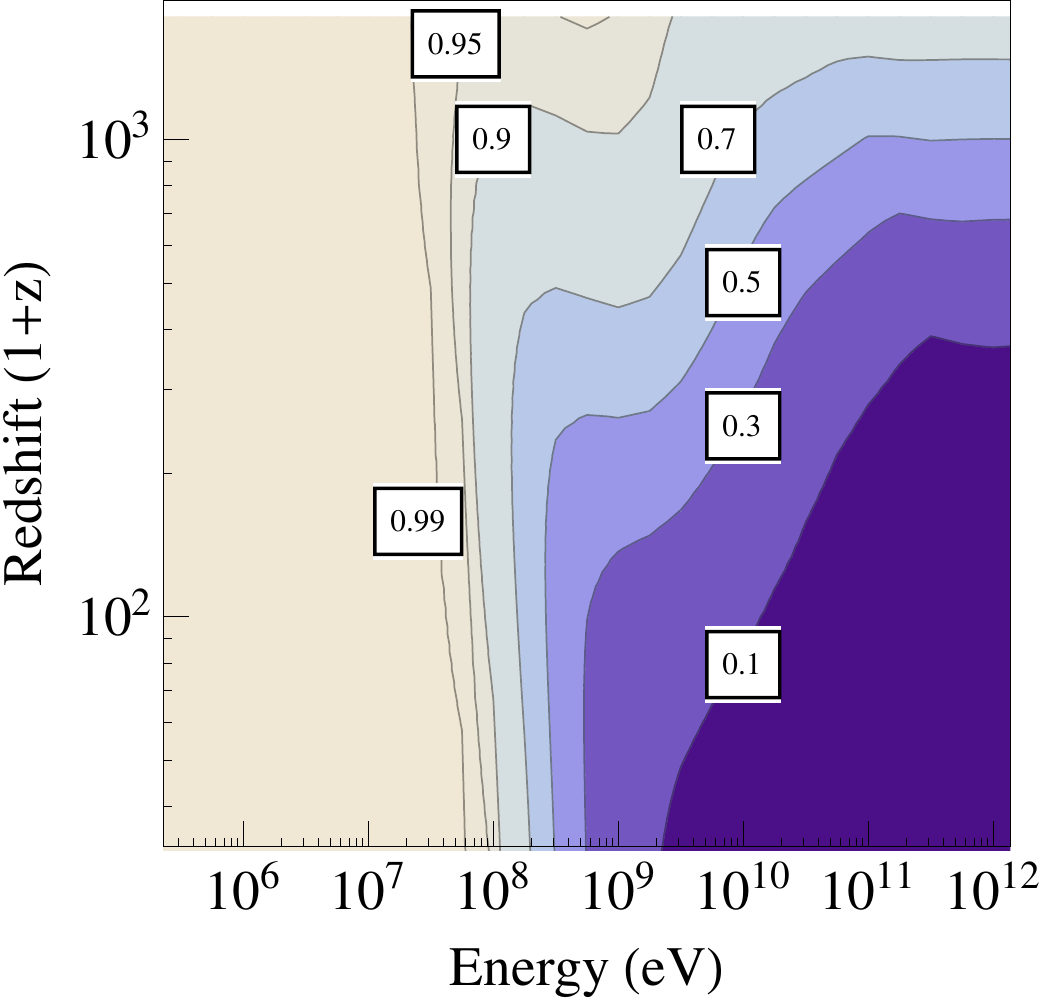}
\caption{\label{fig:totaldeposition}
Fraction of initial energy deposited by $1+z_\mathrm{final}=10$, by a particle or pair injected at a redshift $z$. The different panels show the results for different species. \emph{Left panel}: electron + positron pair (note ``energy'' on the $x$-axis refers to the \emph{kinetic} energy of the electron or positron individually, not the pair -- we assume the electrons and positrons have the same spectrum -- but the plotted values give the fraction of the pair's total (mass + kinetic) energy that is absorbed), \emph{center panel:} photon,  \emph{right panel:} electron (in this case, as there is no way to liberate the electron's mass energy, we plot the fraction of kinetic energy which is absorbed). The energy axis on the right panel is truncated because the behavior is uninteresting at lower energies; when the electron's kinetic energy is small compared to its mass energy, the deposition fraction $\approx 1$.
}
\end{figure*}

As a first step, let us consider the total fraction of the initial particle's energy that is deposited by $z=10$, $p_\mathrm{deposited}$, as a function of initial energy and redshift-of-injection. In terms of our grid elements, this is given by,
\begin{equation} p_\mathrm{deposited}(z_\mathrm{inj}^i, E^k) = \sum_j T_{e^+e^-, e^-, \gamma}^{ijk}. \end{equation}
Fig. \ref{fig:totaldeposition} shows the results for electron-positron pairs, photons, and electrons unaccompanied by a positron.

The fraction of deposited energy falls with increasing redshift, since the universe becomes more transparent as it expands. In general, the fraction of  energy eventually deposited is higher for particles injected at lower energy; this can be attributed to the shorter cooling times for such particles, so more of the energy deposition takes place at higher redshifts where the universe is more opaque. However, electron-positron pairs injected very close to threshold deposit a smaller fraction of their total energy: their \emph{kinetic} energy is efficiently deposited, but their \emph{mass} energy is converted into $\sim 511$ keV photons (or equivalently, the mass energy of the positron and a single ambient electron is so converted), which at low redshifts lie within the ``transparency window'' and can free-stream to the present day.

\subsection{Effective deposition efficiency and comparison with previous results}

Deposition efficiency curves were presented for a range of DM models in \cite{Slatyer:2009yq}. All of these models represent examples of ``conventional'' DM annihilation, in that their annihilation cross section is independent of redshift and so the annihilation rate scales simply as the DM number-density squared. Suppose that, as in this case, the spectrum of the injected particles is redshift-independent and the redshift dependence of the injection rate is known. Then integrating our grid over the redshift-of-injection, weighted by the appropriate redshift-dependent injection rate, yields the total amount of energy deposited at any given redshift.

It is convenient to normalize this energy-deposition history by an energy \emph{injection} history. We will generally normalize to the energy injected from annihilation or decay of the smooth DM component, as appropriate for the model we are studying, since this has a simple redshift dependence. For example, for DM annihilation and ignoring structure formation, the power injected per unit volume at redshift $z$ is given by,
\begin{align} \left(\frac{dE}{dt dV} \right)_\mathrm{injected} & = (1+z)^6 \Omega_\mathrm{DM}^2 c^2 \rho_c^2 \frac{ \langle \sigma v \rangle}{m_\mathrm{DM}}, \label{eq:einjected} \end{align}
where $\langle \sigma v \rangle$ describes the annihilation cross section, $m_\mathrm{DM}$ is the DM mass, and $\Omega_\mathrm{DM}$ is (as usual) the cosmological DM density as a fraction of the critical density $\rho_c$. Following the notation of \cite{Slatyer:2009yq}, we can describe the energy-deposition history by the dimensionless function $f(z)$, defined as,
\begin{align} \left(\frac{dE}{dt dV} \right)_\mathrm{deposited} & = f(z)  \left(\frac{dE}{dt dV} \right)_\mathrm{injected}. \end{align}

The $f(z)$ curves describing energy deposition from DM annihilation (again ignoring structure formation) can readily be extracted from our tables $T^{ijk}$, for any spectrum of photons and $e^+ e^-$ pairs produced by the annihilation. Furthermore, it is straightforward to include cases where the spectrum of annihilation products is redshift-dependent (for example, in the case where there are multiple annihilation channels with different velocity or time dependences).

Defining the interpolating function $T(z_\mathrm{inj},z_\mathrm{dep},E)$ as above, so $T^{ijk} = T(z^i_\mathrm{inj},z^j_\mathrm{dep},E^k) d \ln (1 + z_\mathrm{dep})$, we can write,
\begin{align} f(z) & = \frac{\displaystyle \sum_\mathrm{species} \int E dE \int d z' T(z',z,E) \frac{dN}{dE d z'}}{\displaystyle \sum_\mathrm{species} \int E \left(\frac{dN}{dE d\ln (1 + z)} \right)_\mathrm{norm} dE }, \label{eq:effectivef} \end{align}
where $\frac{dN}{dE d\ln(1+z)}$ describes the spectrum of injected particles (of a particular species), per comoving volume, as a function of energy and redshift. $\left(\frac{dN}{dE d\ln(1+z)}\right)_\mathrm{norm}$ refers to the energy injection history we are using for normalization (we will typically pick something simple, like annihilation or decay of DM neglecting structure formation). 

For example, studying energy deposition from conventional DM annihilation, where the energy injection histories can be factored into a redshift dependence and a spectrum per annihilation $d\bar{N}/dE$, it is natural (and agrees with the convention in  \cite{Slatyer:2009yq}) to normalize to the injected energy from DM annihilation in the same model, thus canceling out redshift-independent, energy-independent proportionality factors in Eq. \ref{eq:effectivef} such as  $\langle \sigma v \rangle$. For conventional DM annihilation,
\begin{align} \frac{dN}{dE d\ln (1+z)} & = \left(\frac{dN}{dE dV dt} \right) dV \frac{dt}{d\ln(1+z)} \nonumber \\ & \propto  \left(\frac{d\bar{N}}{dE} \right) \frac{(1+z)^3}{H(z)},  \end{align}
using $\frac{dN}{dE dV dt} \propto (1+z)^6$, $dV \propto (1+z)^3$ and $H(z) = - \frac{d \ln (1+z)}{dt}$, thus yielding,
\begin{align} f(z) & = \frac{H(z)}{(1+z)^3 \displaystyle \sum_\mathrm{species} \int E \frac{d\bar{N}}{dE} dE} \times \nonumber \\ & \displaystyle \sum_\mathrm{species} \int \frac{(1+z')^2 dz'}{H(z')} \int T(z',z,E) E \frac{d\bar{N}}{dE} dE. \label{eq:annf} \end{align}
For conventional DM decay, with a lifetime much longer than the age of the universe (i.e. $\frac{dN}{dE dV dt} \propto (1+z)^3$) and a spectrum-per-decay of $d\bar{N}/dE$, the corresponding expression is,
\begin{align} f(z) & = H(z) \frac{\displaystyle \sum_\mathrm{species} \int \frac{d\ln(1 + z')}{H(z')} \int T(z',z,E) E \frac{d\bar{N}}{dE} dE }{\displaystyle \sum_\mathrm{species} \int E \frac{d\bar{N}}{dE} dE}. \end{align}
Here we have normalized the energy deposition curve to the energy injection curve from the same process; in both cases we can therefore think of $f(z)$ as an ``effective efficiency'' mapping injection to deposition.

\begin{figure*}
\includegraphics[width=.45\textwidth]{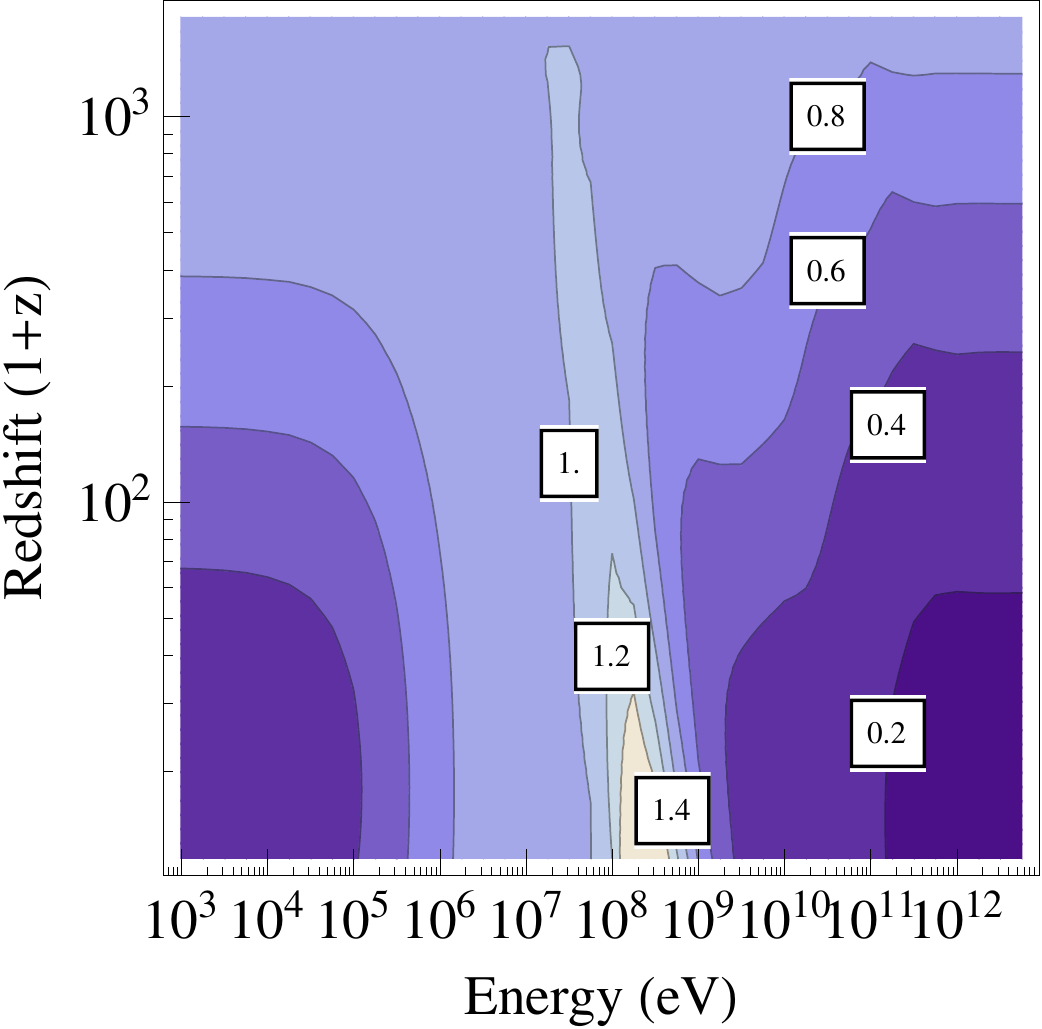}
\includegraphics[width=.45\textwidth]{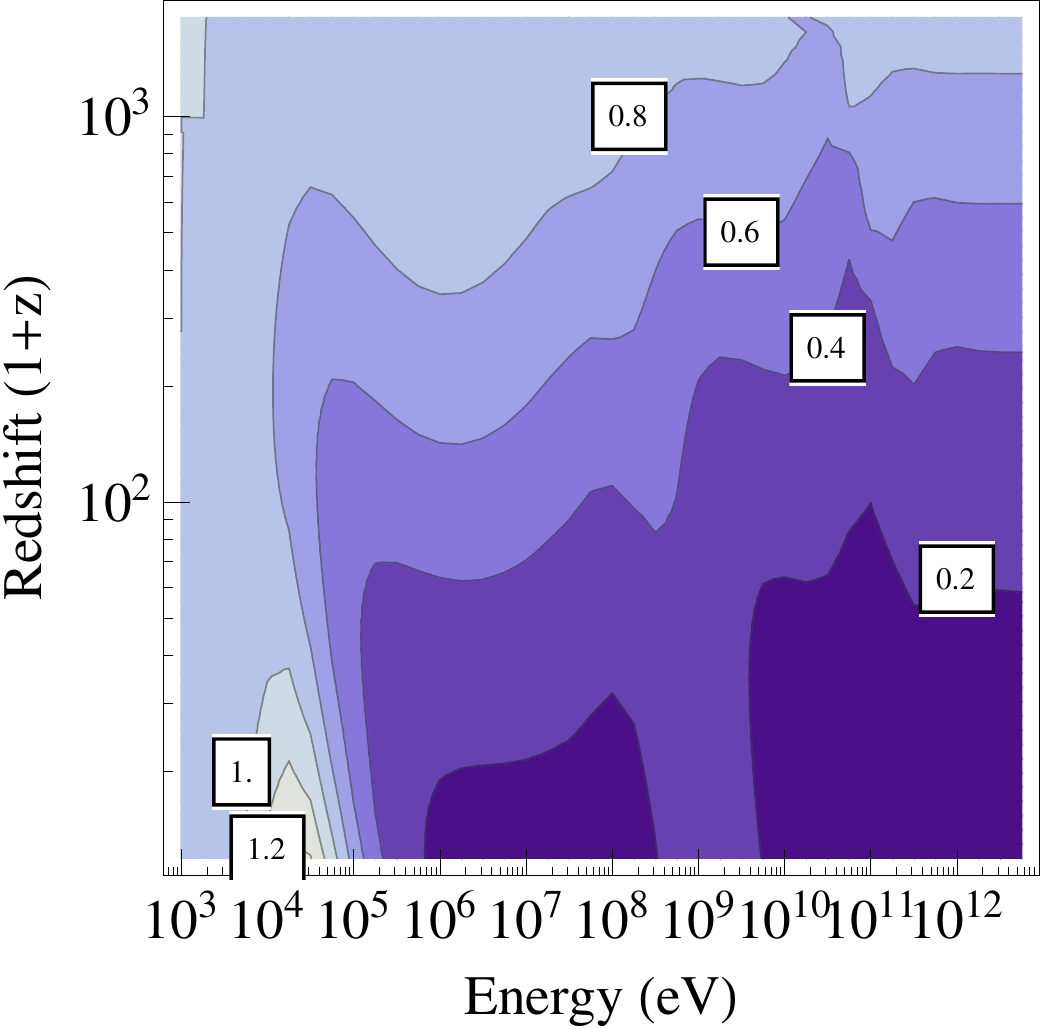}  \\
\includegraphics[width=.45\textwidth]{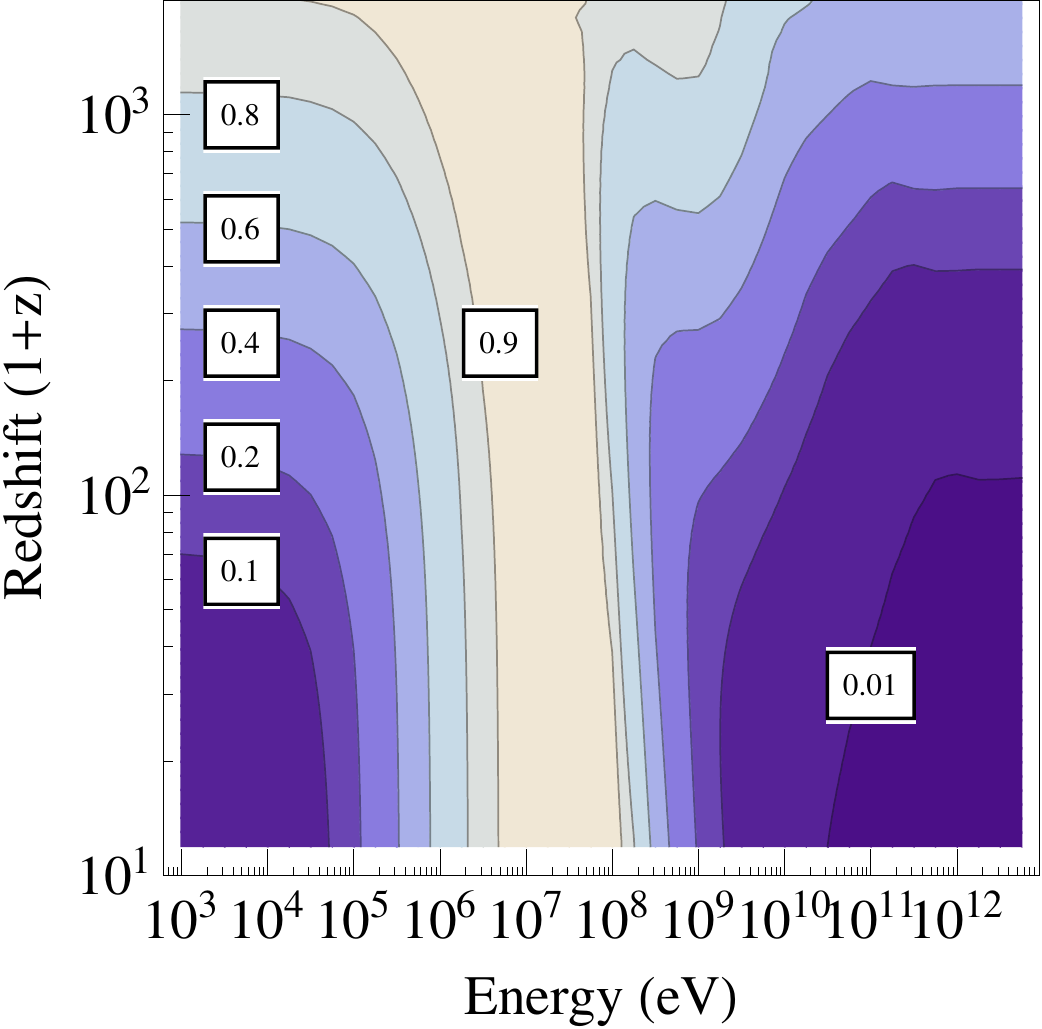}
\includegraphics[width=.45\textwidth]{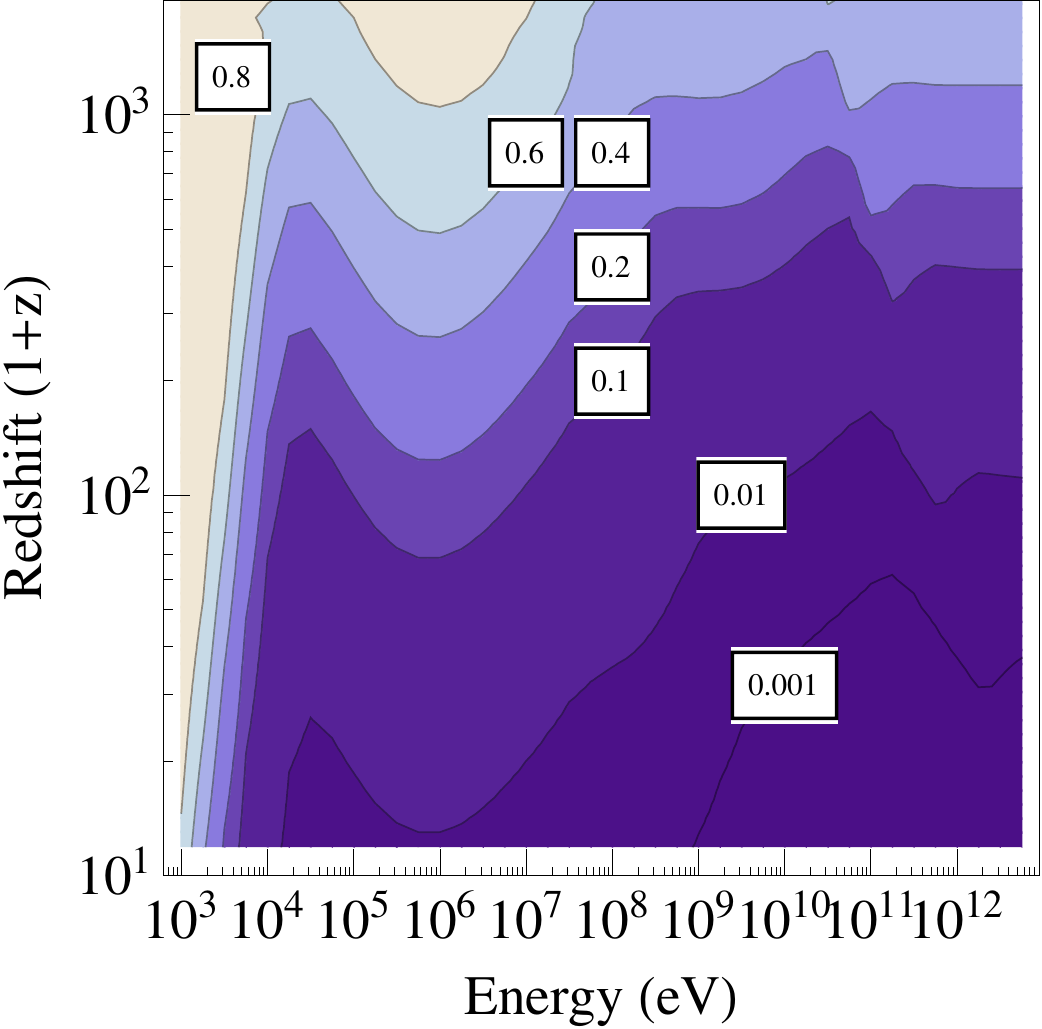}
\caption{\label{fig:fDM}
Effective efficiency (i.e. the ratio of the energy deposition history to the energy injection history) for (\emph{top}) annihilating DM and (\emph{bottom}) decaying DM, for electron-positron pairs (\emph{left-hand panels}) and photons (\emph{right-hand panels}), as a function of redshift-of-deposition and initial energy of the photon/electron/positron.
}
\end{figure*}

Fig. \ref{fig:fDM} shows the efficiency curves for photons and $e^+e^-$ pairs as a function of injection energy and redshift-of-deposition, for injection histories corresponding to annihilating and decaying DM. In the annihilation case, we have included only energy injection corresponding to the smooth DM relic density, neglecting the onset of structure formation below $z \sim 100$.

Note that unlike the total fraction of energy deposited, the effective efficiency can rise substantially at low redshifts, particularly in the annihilating-DM case. The increased transparency of the universe is more than compensated by the residual pool of photons produced by more-rapid energy injection at high redshifts, which remain in the transparency window for a long period and only deposit the bulk of their energy at lower redshifts. The efficiency is systematically higher for annihilating DM than decaying DM, because annihilating DM injects more power at higher redshifts when the universe is more opaque.

\begin{figure*}
\includegraphics[width=.45\textwidth]{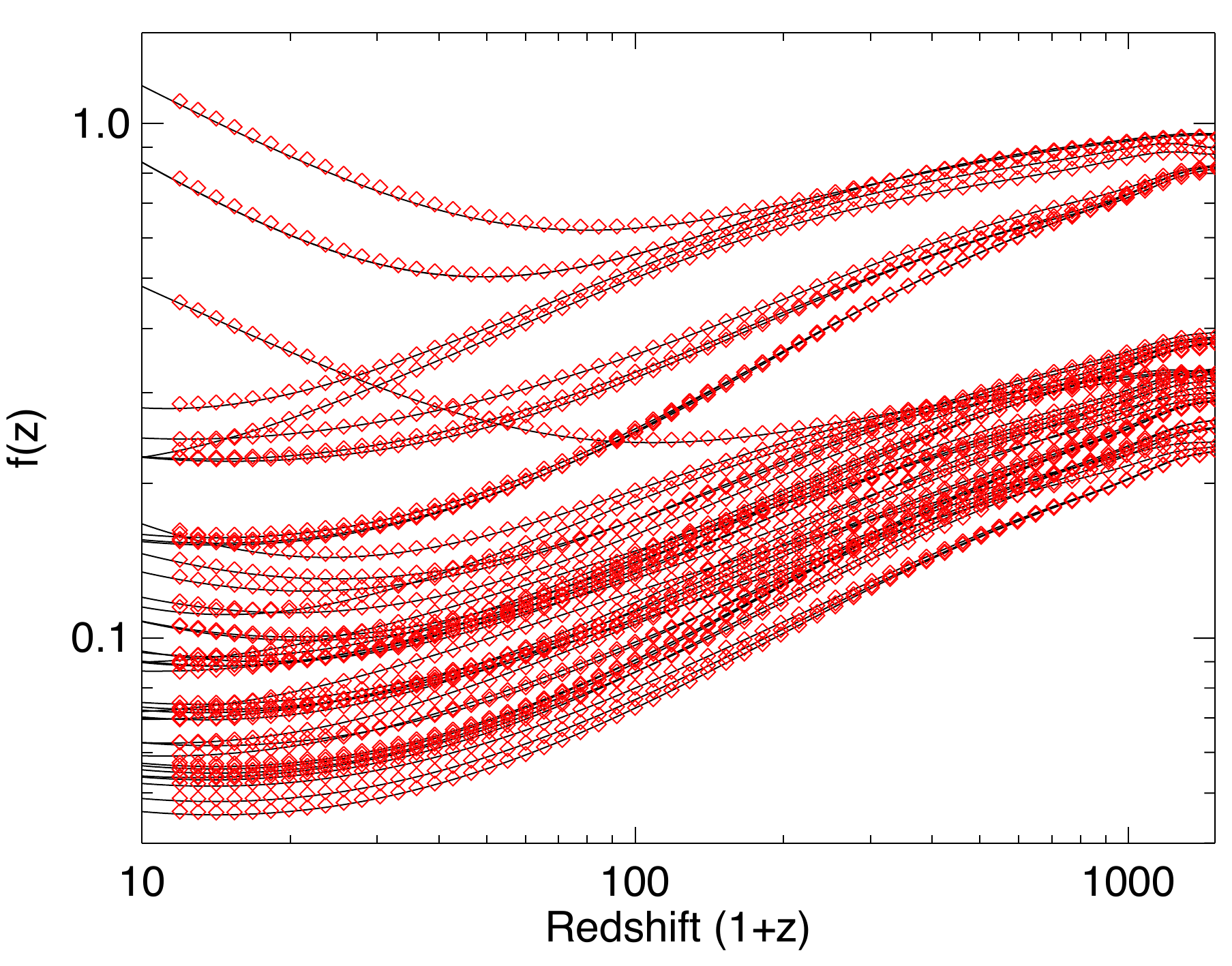}
\includegraphics[width=.45\textwidth]{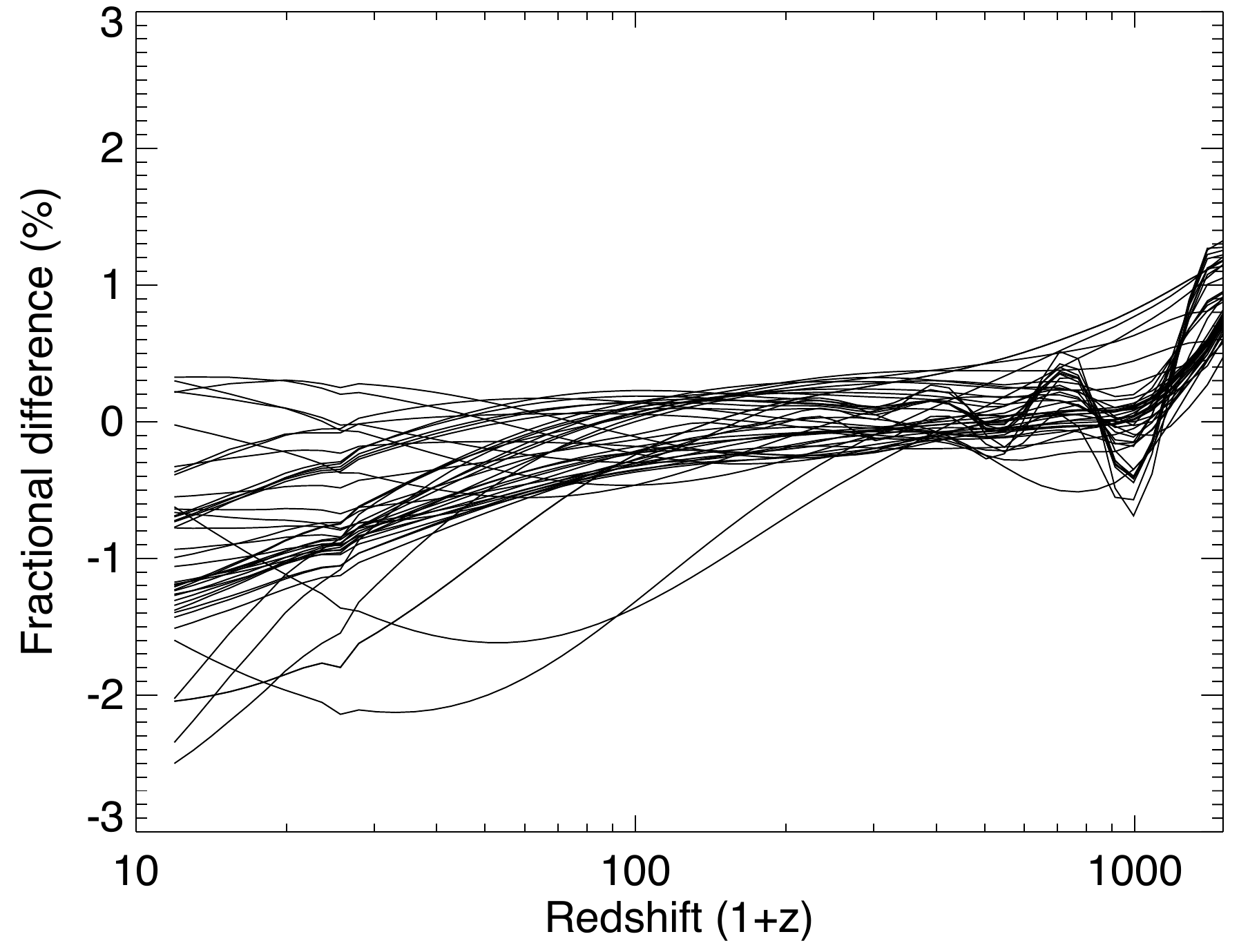}
\caption{\label{fig:comparison}
\emph{Left panel:} The $f(z)$ curves derived in \cite{Slatyer:2009yq} for 41 models of interest (solid black lines), and the recovered $f(z)$ curves using the results presented in this paper (red diamonds), between $z=10$ and $z=1500$. \emph{Right panel:} The percentile difference between the two curves, between $z=10$ and $z=1500$.  
}
\end{figure*}

In Fig. \ref{fig:comparison} we compare the effective-efficiency curves obtained by this procedure to those given in \cite{Slatyer:2009yq} for annihilating DM, for the same SM final states and DM masses listed there. For each final state and DM mass we take the injected spectrum of photons and $e^+ e^-$ pairs calculated using \texttt{Pythia} (as in \cite{Slatyer:2009yq}), integrate these spectra over the energy- and species-dependent effective-efficiency curves plotted in Fig. \ref{fig:fDM}, and compare to the $f(z)$ curves from \cite{Slatyer:2009yq}. We obtain excellent agreement, indicating that our rather coarse binning is nonetheless adequate for the purpose.

\subsection{Inclusion of structure formation}

Constraints on DM annihilation from the CMB are typically computed only taking into account the smooth component of the DM. This is appropriate at high redshift, when density fluctuations are small and perturbative, and conservative in all cases; it allows a very robust and precise constraint, since the cosmological density of the DM is constrained at the percent level. However, there has been considerable interest in the effects of DM annihilation/decay at lower redshifts ($z \lesssim 30$) where there might be measurable effects on the 21 cm line of hydrogen \cite{Furlanetto:2006wp, Valdes:2007cu,Cumberbatch:2008rh,Finkbeiner:2008gw,Natarajan:2009bm,Valdes:2012zv}, or where the annihilation byproducts might help reionize the universe \cite{Natarajan:2008pk, Belikov:2009qx, Natarajan:2010dc,Giesen:2012rp, Natarajan:2012ry}.

At these redshifts it becomes crucial to account for the contribution to annihilation from bound DM halos; since the annihilation rate scales as the square of the density, annihilation in halos eventually greatly dominates over the smooth component. Some previous studies have assumed an on-the-spot approximation, taking the energy deposition rate to be equal to the energy injection rate or making some simple ansatz for the fraction of injected energy that is promptly absorbed; others have multiplied the energy deposition rate (e.g. parametrized by $f(z)$ as defined above) by a factor describing the extra annihilation from halos at that redshift. However, these approaches can give rather inaccurate results, because at low redshifts the absorption of annihilation products' energy is often significantly delayed; the annihilation rate in bound halos that is relevant is set by the redshift of \emph{injection}, not deposition.

There are substantial uncertainties on the contribution to the total annihilation rate from bound subhalos -- the smallest halos are expected to be far below the resolution of $N$-body simulations -- but in the present context we will use the simple prescription outlined in \cite{Giesen:2012rp} to demonstrate the effect of delayed energy deposition. Employing the Press-Schechter differential mass function, the authors of \cite{Giesen:2012rp} find that the energy injection rate from annihilation in halos can be parameterized by,
\begin{equation} \frac{dE}{dV dt} = \rho_c^2 \Omega_\mathrm{DM}^2 c^2 (1+z)^3 \mathrm{erfc}\left(\frac{1+z}{1+z_h} \right) \bar{f}_h \frac{\langle \sigma v \rangle}{m_\mathrm{DM}}, \end{equation}
where the redshift of halo formation $z_h$ and the normalization factor $\bar{f}_h = \frac{200}{3} (1+z_F)^3 f_\mathrm{NFW}(c_h)$ should be extracted from $N$-body structure formation simulations. \cite{Giesen:2012rp} considers $z_h \sim 20$ and $\bar{f}_h \sim 10^9 - 10^{10}$ to be plausible values for these parameters. 

Factoring out the energy injection rate from the smooth DM density (Eq. \ref{eq:einjected}), the enhancement factor from including halos is given by,
\begin{equation} 1 + \frac{\bar{f}_h}{(1+z)^3}  \mathrm{erfc}\left(\frac{1+z}{1+z_h} \right). \end{equation}

\begin{figure*}
\includegraphics[width=.45\textwidth]{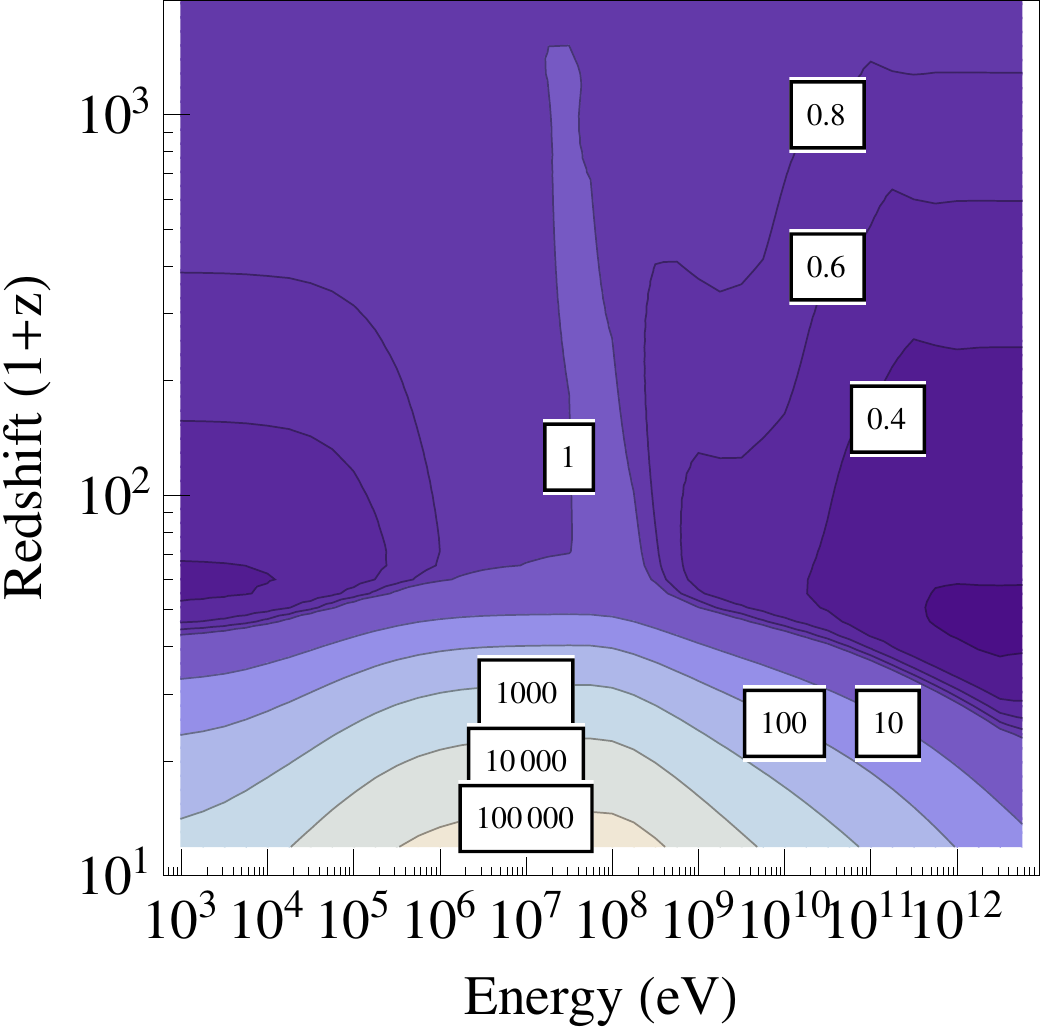}
\includegraphics[width=.45\textwidth]{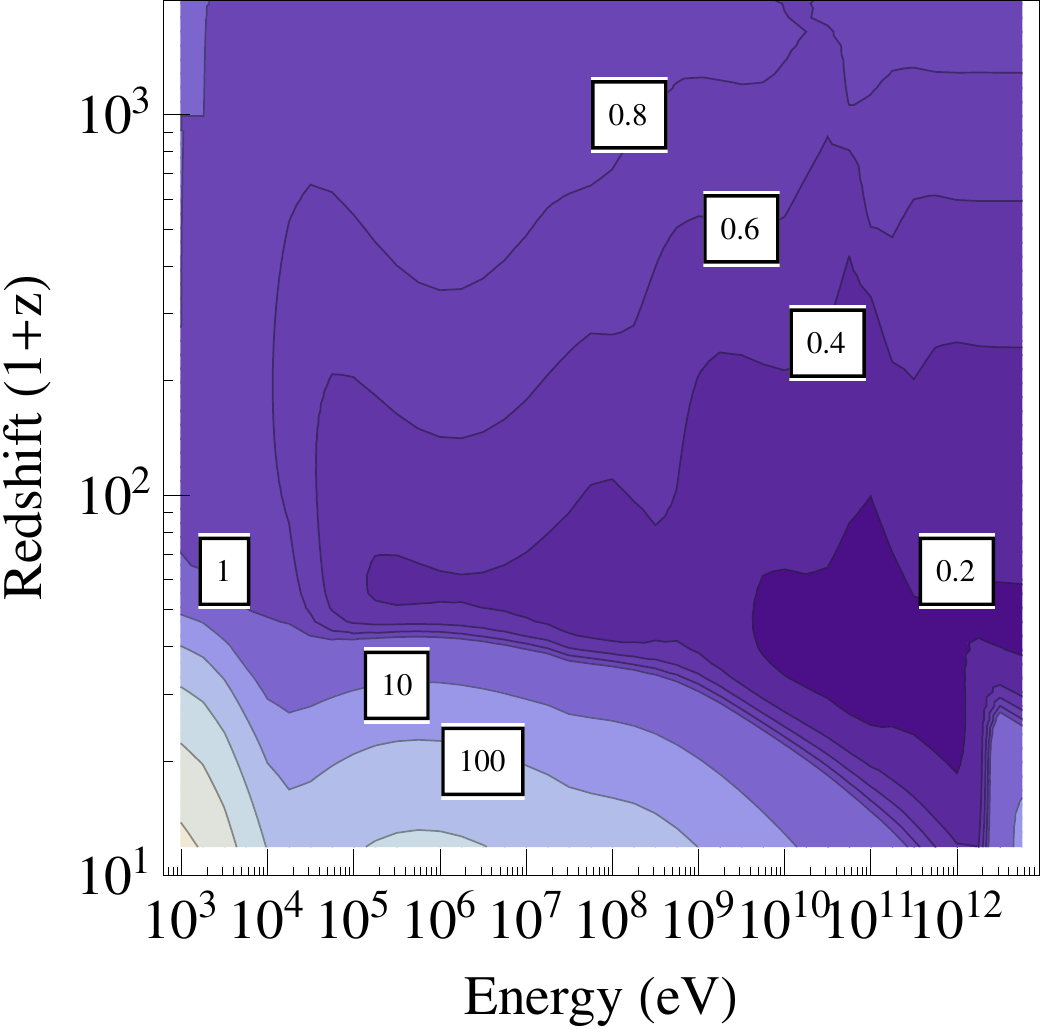}
\caption{\label{fig:structureformation}
The ratio of the energy deposited, from the halos + the smooth component, to the energy injected from the smooth component, as a function of injection energy and redshift-of-deposition, for (\emph{left panel}) $e^+ e^-$ pairs, (\emph{right panel})  photons.
}
\end{figure*}

\begin{figure}
\includegraphics[width=.48\textwidth]{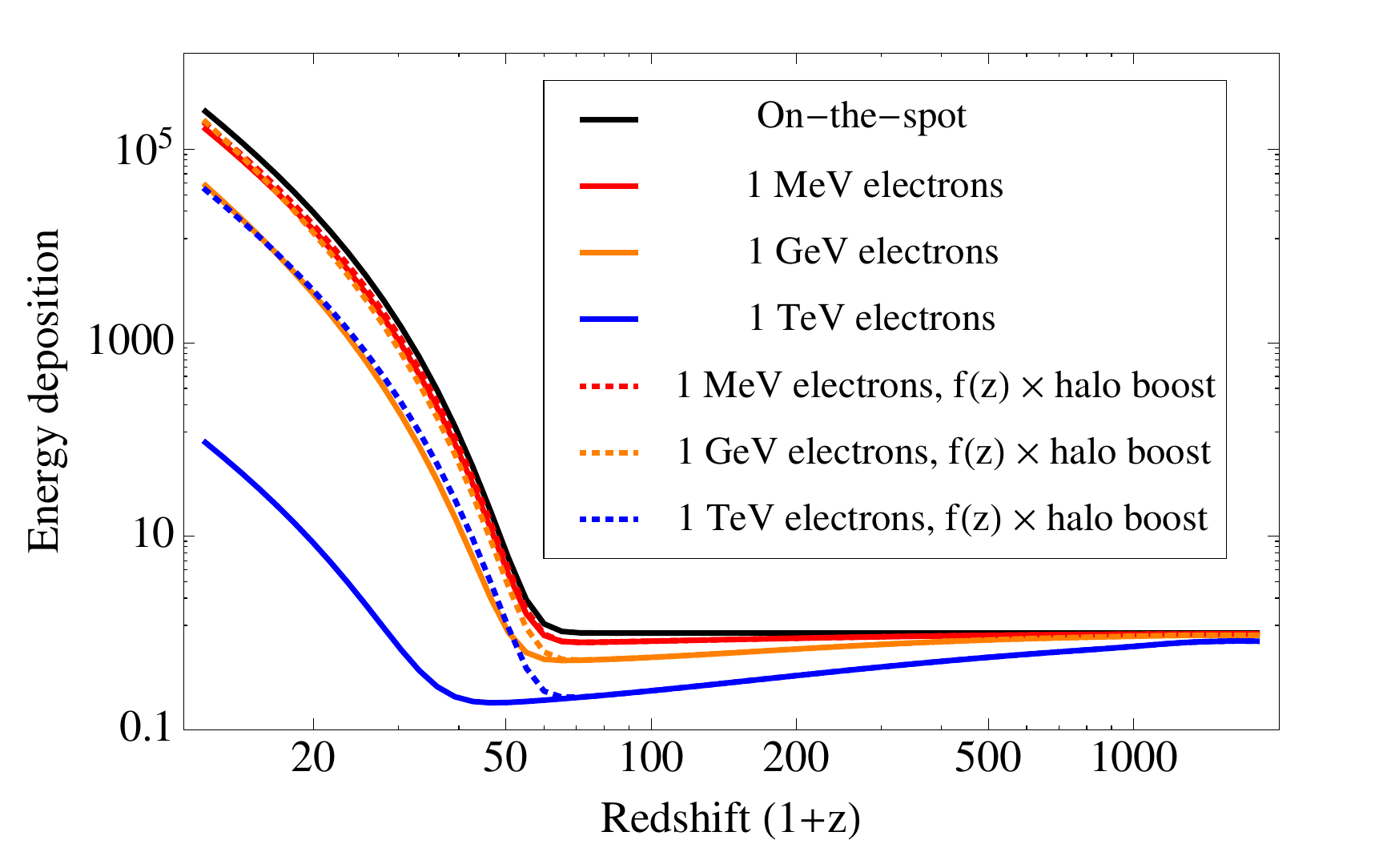}
\caption{\label{fig:onthespotcompare}
The ratio of the energy deposited, from the halos + the smooth component, to the energy injected from the smooth component, as a function of redshift-of-deposition, for three different electron injection energies (assuming they are injected as a pair with a positron of the same energy), as well as the on-the-spot curve where the deposited energy at every redshift is equal to the injected energy (from halos + smooth component). Dotted lines show the result of simply multiplying the $f(z)$ derived for the smooth component by the enhancement factor from the halos, rather than correctly inserting the enhancement factor \emph{inside} the integral in Eq. \ref{eq:effectivef}.
}
\end{figure}

Given any ansatz for the boost to DM annihilation from halos, since we have the full function mapping injection to deposition, it is trivial to compute the modified energy deposition curve: we simply multiply the spectrum of injected particles, $\frac{dN}{dE d\ln(1+z)} \propto  \left(\frac{d\bar{N}}{dE} \right) \frac{(1+z)^3}{H(z)}$, by the appropriate redshift-dependent halo enhancement, \emph{inside} the integral in Eq. \ref{eq:effectivef}. Fig. \ref{fig:structureformation} shows  the dimensionless energy deposition function $f(z)$ as a function of redshift with the halo parameterization above, taking $z_h = 20$ and $\bar{f}_h = 10^9$, normalized to the energy injection at that redshift from the \emph{smooth component only}. (This removes the interpretation of $f(z)$ as an effective efficiency, but facilitates comparisons with the baseline case where we ignore structure formation.) For comparison, in Fig. \ref{fig:onthespotcompare} we  show the difference between these results and an on-the-spot approximation, taking all the injected energy to be promptly deposited (i.e. $f(z) = 1 + \frac{\bar{f}_h}{(1+z)^3}  \mathrm{erfc}\left(\frac{1+z}{1+z_h} \right)$). We see that the effects of delayed energy deposition can be substantial, especially for higher-mass DM. We also show the results of the common approximation of simply multiplying the $f(z)$ derived for the smooth component by the enhancement factor from the halos; again, this is a rather poor approximation for high-mass DM.

\section{CMB limits on unconventional injection histories: a review}
\label{sec:review}

It is straightforward to obtain deposition histories for models with unconventional redshift-dependent energy injection histories, simply by changing the factor by which we multiply the grid when summing over the redshift-of-injection. Given such a deposition history, we can immediately apply the principal-component methods of \cite{Finkbeiner:2011dx} to estimate the CMB constraints on such a class of models\footnote{These methods include an approximate formulation for the partition of ``deposited'' energy between ionization, excitation, heating and low-energy (CMB-like) photons for which the universe is transparent; these approximations may mis-estimate the bounds by several tens of percent \cite{Evoli:2012qh}.}. For convenience, we briefly review the method here.

Suppose we write the energy deposition history $\frac{dE}{dV dt}$ as some dimensionless function $f(z)$, multiplying a dimensionful normalization factor $\varepsilon$ and a fixed baseline energy injection history with no dependence on the model under study, $b(z)$. For example, for energy injection from DM annihilation with DM mass $m_\chi$ and annihilation cross section $\langle \sigma v \rangle$, we typically choose,
\begin{equation} b(z) = (1+z)^6 \Omega_\mathrm{DM}^2 c^2 \rho_c^2, \quad \varepsilon = \frac{ \langle \sigma v \rangle}{m_\mathrm{DM}},\end{equation}
and for DM decay with lifetime $\tau$,
\begin{equation} b(z) = (1+z)^3 \Omega_\mathrm{DM} \rho_c c^2, \quad \varepsilon = \frac{1}{\tau}.\end{equation}

We can now parameterize the redshift dependence of the energy deposition, described by $f(z)$, by its coefficients in a basis of orthonormal principal components $\{ e_i \}$ supplied by \cite{Finkbeiner:2011dx},
\[ \varepsilon f(z) = \sum_{i=1}^N \varepsilon_i e_i(z), \quad \varepsilon_i = \varepsilon \frac{f(z) \cdot e_i(z)}{e_i(z) \cdot e_i(z)} = \varepsilon f(z) \cdot e_i(z), \]
where the dot product is defined appropriately as a sum over discrete variables or an integral over a product of functions. Principal components have been derived by  \cite{Finkbeiner:2011dx} for both choices of $b(z)$ described above; of course, this is just a choice of basis, but the appropriate choice of basis means the effect of the energy injection on the CMB can be adequately described by a smaller number of principal components.

Each principal component $e_i(z)$ has an eigenvalue $\lambda_i$ associated with it (with different eigenvalues for \WMAP7, \PLANCKc, a cosmic variance limited (CVL) experiment, etc), which describes its detectability when multiplied by some canonical (dimensionful) normalization factor $\varepsilon_i = \bar{\varepsilon} \, \forall \, i$. The total significance of the signal is then given (in the Gaussian approximation) by $\sqrt{\sum_i \lambda_i \varepsilon_i^2} / \bar{\varepsilon}$, so for example, the $2 \sigma$ limit corresponds approximately to the constraint,
\[\varepsilon < \frac{2 \bar{\varepsilon}}{\sqrt{\sum_i \lambda_i \left( f \cdot e_i\right)^2}}.\]
Alternatively, $\bar{\varepsilon}/\sqrt{\lambda_i}$ can be replaced by the error bar on energy injection corresponding to the $i$th PC, $\sigma_i$, as obtained from a likelihood analysis using \texttt{CosmoMC}, in which case the limit is,
\[\varepsilon < \frac{2}{\sqrt{\sum_i \left( \frac{f \cdot e_i }{\sigma_i}\right)^2}}.\]
For \PLANCK forecasting, the two limits are essentially identical, as the approximations used to derive the eigenvalues are very good for the energy injections constrained by \PLANCK. For energy injections at the \WMAP7 limit, one of the approximations begins to break down; more accurate results will be obtained by using the \texttt{CosmoMC} bound on the first principal component, that is,
\[ \varepsilon < \frac{1.2 \times 10^{-26} \mathrm{cm}^3\mathrm{/s/GeV}}{f \cdot e_1} \]
Inclusion of ACT data tends to strengthen constraints from \WMAP7 by $\sim 15\%$ \cite{Galli:2011rz}.

For \PLANCK and a CVL experiment, where including more than one principal component can be justified (see \cite{Finkbeiner:2011dx} for details), the values of the first few $\sigma_i$ are,
\begin{itemize}
\item \PLANCKc: 

$\{ \sigma_1, \sigma_2, \sigma_3 \} = \{1.1, 2.4, 4.1\} \times 10^{-27} \mathrm{cm}^3\mathrm{/s/GeV}, $
\item CVL: 

$\{ \sigma_1, \sigma_2, \sigma_3, \sigma_4, \sigma_5 \} = \{0.5, 1.1, 1.8, 2.5, 3.4\} \times 10^{-27} \mathrm{cm}^3\mathrm{/s/GeV}. $
\end{itemize}

Alternative principal components were derived for decaying species (the second $b(z)$ profile given above), since decay-like energy injection histories have much greater relative effects at low redshift than annihilation-like energy injection histories. The principles are identical, but a likelihood analysis with \texttt{CosmoMC} has not been performed. The values of the first few $\bar{\varepsilon}/\sqrt{\lambda_i}$ are:
\begin{itemize}
\item \WMAPc: 

$\{ \sigma_1, \sigma_2, \sigma_3 \} = \{5.3, 6.4, 9.5\} \times 10^{-25} \mathrm{s}^{-1}, $

\item \PLANCKc: 

$\{ \sigma_1, \sigma_2, \sigma_3 \} = \{1.2, 1.6, 2.3\} \times 10^{-25} \mathrm{s}^{-1}, $
\item CVL: 

$\{ \sigma_1, \sigma_2, \sigma_3, \sigma_4, \sigma_5 \} = \{2.6, 5.7, 8.0, 12, 20\} \times 10^{-26} \mathrm{s}^{-1}. $
\end{itemize}

\section{New and Updated Constraints}
\label{sec:constraints}

\subsection{Limits on late-decaying species}

DM itself may decay, with a lifetime much longer than the age of the universe; alternatively, new metastable states might decay, either to a lighter DM state + Standard Model particles (e.g. \cite{Finkbeiner:2007kk,Bell:2010qt}), or purely into Standard Model particles (in this latter case, the metastable species could only constitute a very small fraction of the matter density). If photons, electrons or positrons are produced in these decays, they can be constrained by the methods discussed here. Such constraints were worked out for the 1-year and 3-year \WMAP data in \cite{Chen:2003gz, Zhang:2007zzh}, employing an on-the-spot approximation.

We consider a decaying species with lifetime $\tau$. Then the rate of energy injection per unit volume is given by,
\begin{align} \left(\frac{dE}{dt dV} \right)_\mathrm{injected} & = (1+z)^3 \Delta \Omega_\mathrm{dec}  \rho_c c^2 \frac{e^{-t/\tau}}{\tau}, \end{align}
where $\Omega_\mathrm{dec}$ is the density of the decaying species (in units of the critical density), and $\Delta$ is the fraction of its mass energy liberated in each decay.

\begin{figure*}
\includegraphics[width=.45\textwidth]{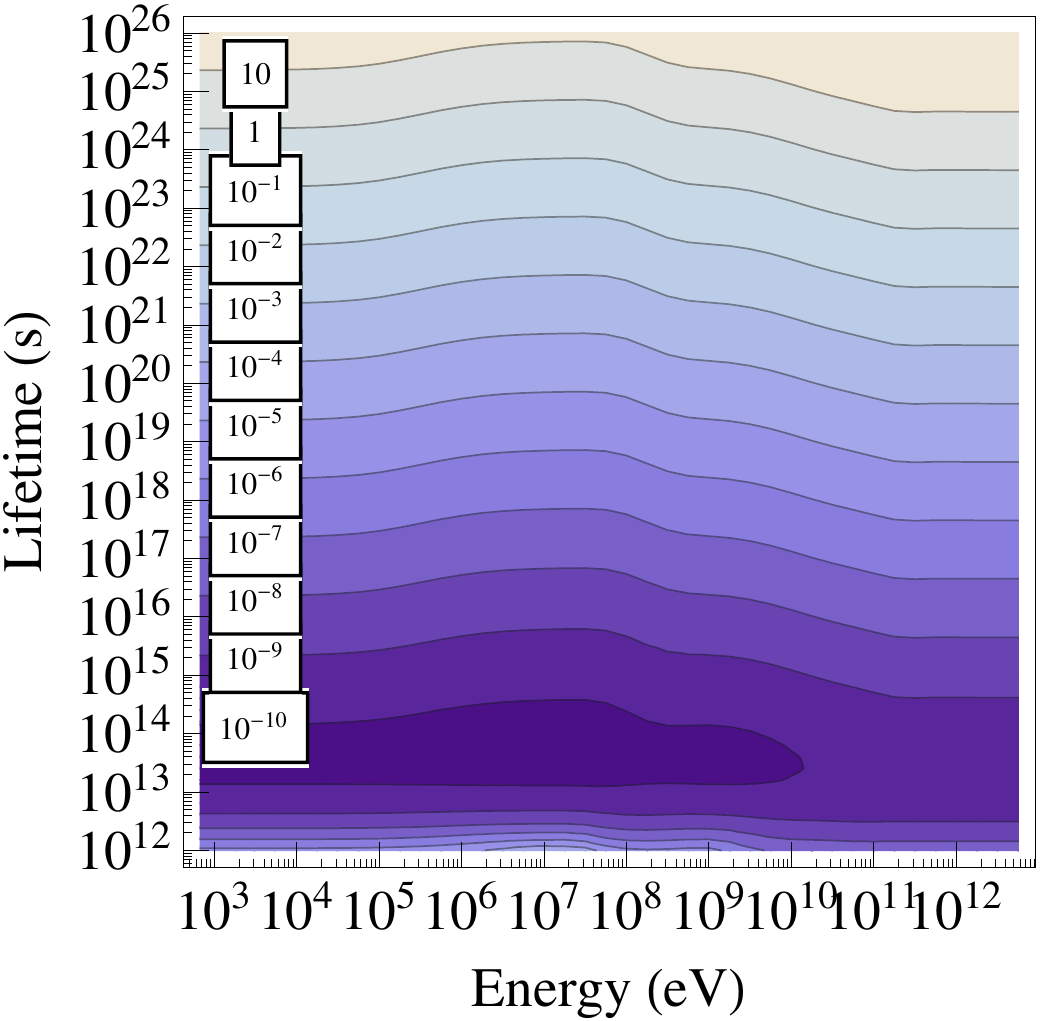}
\includegraphics[width=.45\textwidth]{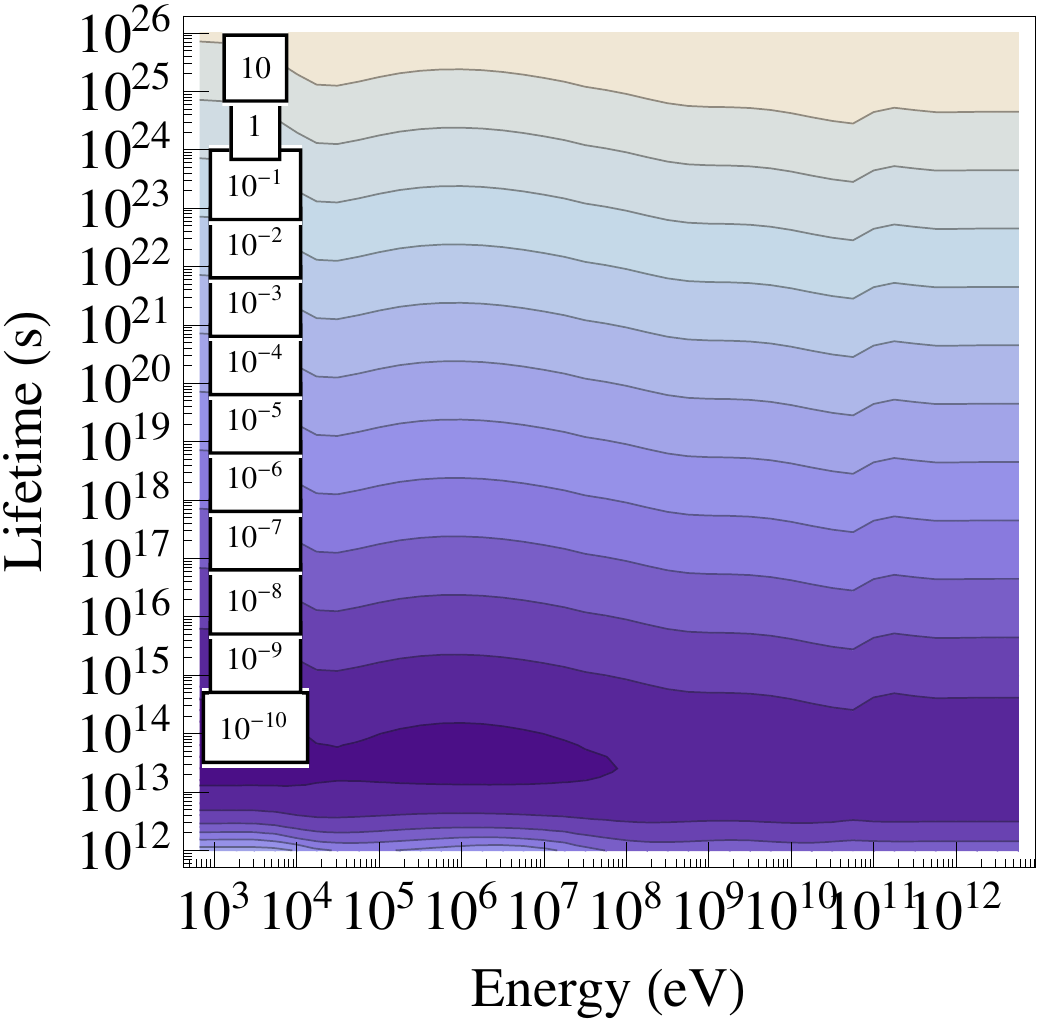} \\
\includegraphics[width=.45\textwidth]{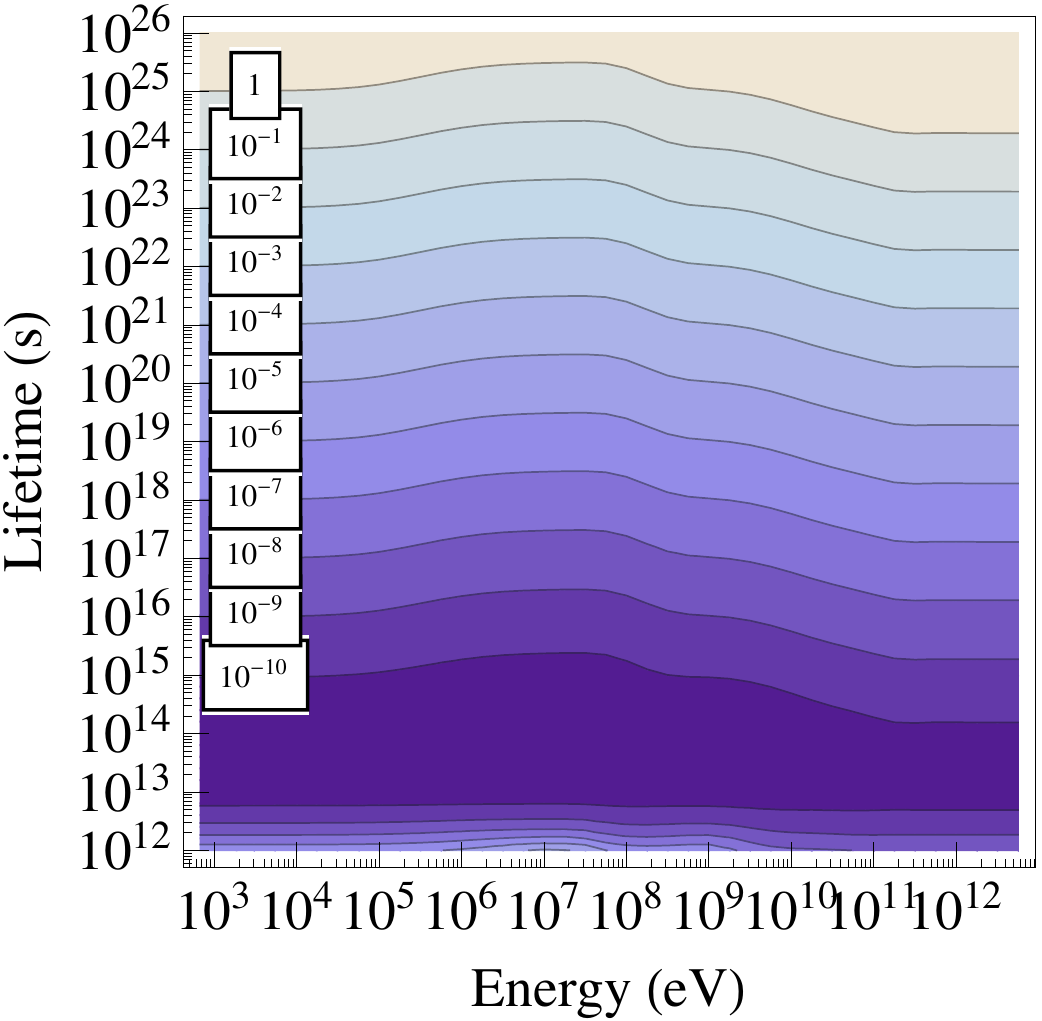}
\includegraphics[width=.45\textwidth]{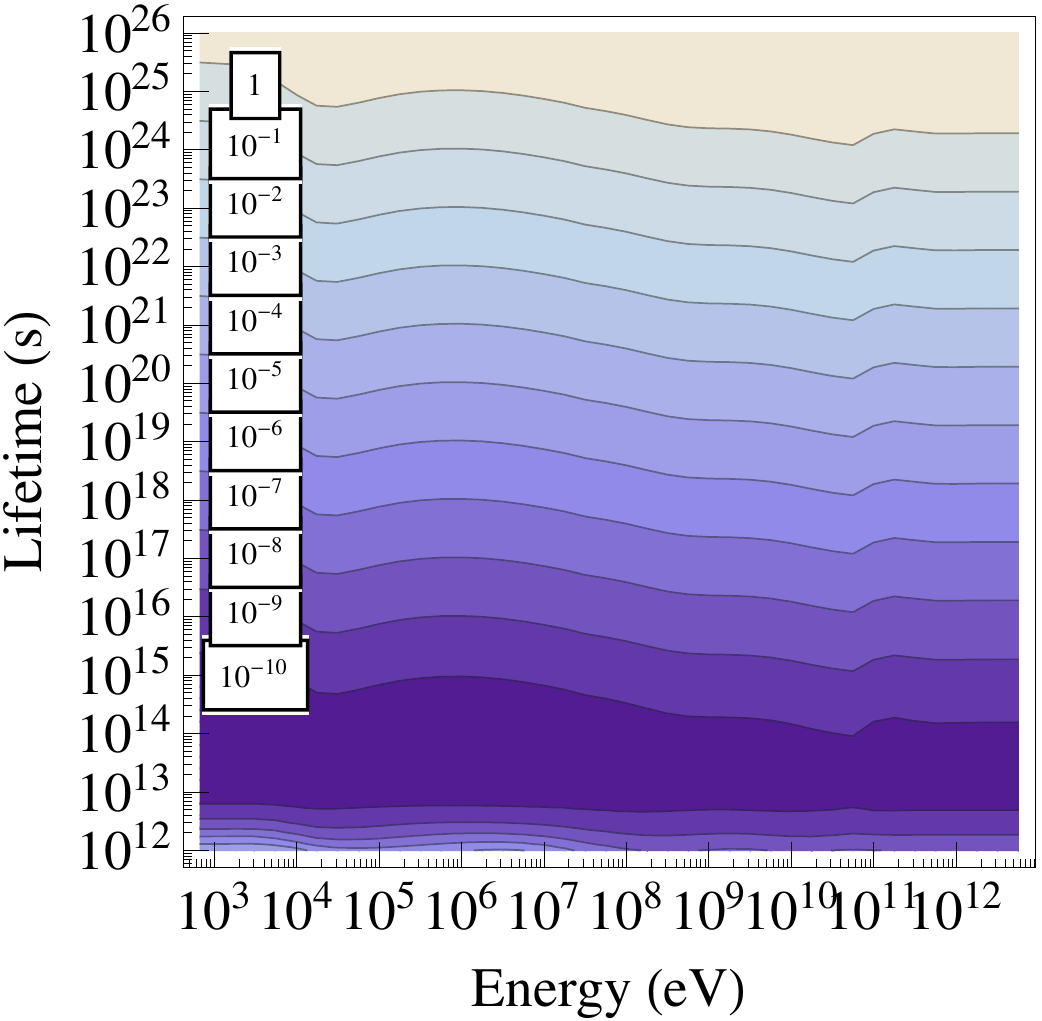}
\caption{\label{fig:decaylimits}
Upper bounds on the fraction of the total DM density that can be composed of a decaying species, as a function of the decay lifetime and the energy of the decay products: the $95\%$ confidence limit from \WMAP7 (\emph{top row}) and the forecast $95\%$ confidence limit from \emph{Planck} (\emph{bottom row}). The left-hand panels apply to $e^+e^-$ pairs (as usual, the energy is for a single particle), the right-hand panels to photons. Note that these constraints also apply to the decay of a metastable excited state; see the text for details.}
\end{figure*}

\begin{figure}
\includegraphics[width=0.45\textwidth]{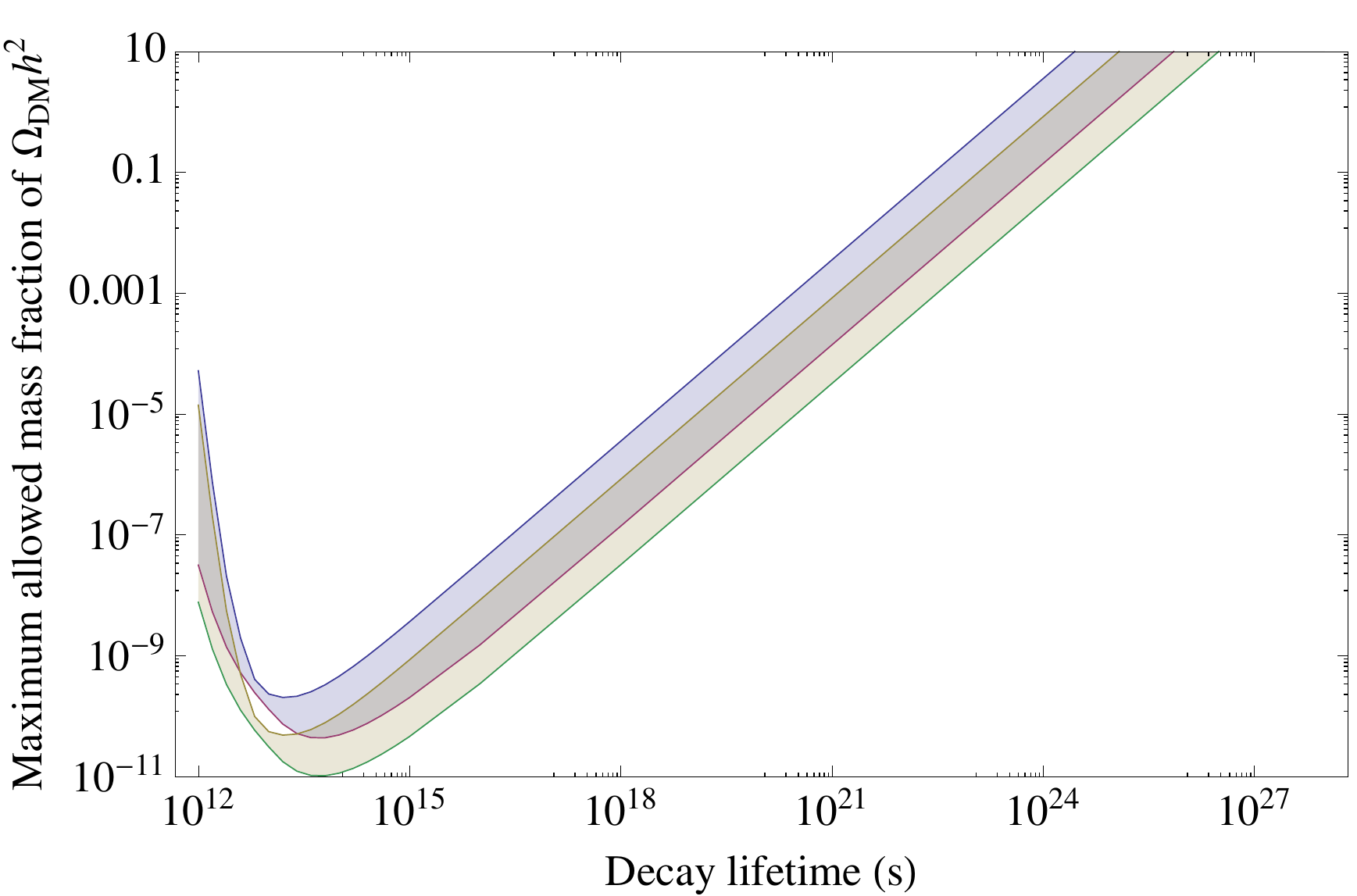}
\caption{\label{fig:decaylimitband}
The $95\%$ confidence limits from \WMAP7 and \emph{Planck} (forecast) on the fraction of the total DM density that can be composed of a decaying species, as a function of the decay lifetime, for a range of decay products and energies. For each lifetime we scan over the constraint for photons and $e^+ e^-$ pairs with (single-particle) energy ranging from 1 keV to 6 TeV, to obtain the bands displayed; the width of the bands reflect the variation in the constraint between different energies and decay products. The upper (blue) band corresponds to \WMAP7 limits, the lower (yellow) band to \emph{Planck}. Note that these constraints also apply to the decay of a metastable excited state; see the text for details. A branching fraction less than unity to electrons/positrons/photons will shift these bands upward. Values above the lines are excluded.
}
\end{figure}

As previously, we can normalize the energy deposition curve to the profile of energy injection for DM decaying with a long lifetime, $b(z) = (1+z)^3 \Omega_\mathrm{DM} \rho_c c^2$; specifically, we define $f(z)$ by,
\begin{equation} \left(\frac{dE}{dt dV} \right)_\mathrm{deposited} = f(z) \left( \frac{\Delta \Omega_\mathrm{dec}}{\Omega_\mathrm{DM}} \frac{1}{\tau} \right) b(z), \end{equation} 
and compute $f(z)$ by summation over the $T^{ijk}$ grid, for any spectrum of decay products and lifetime $\tau$. (Note that now, if deposition precisely traced injection, $f(z)$ would be equal to $e^{-t/\tau}$, not 1.)
For any value of $\tau$, we can then constrain $\Delta \Omega_\mathrm{dec} / \Omega_\mathrm{DM}$ by the CMB as described above, by taking dot products of $f(z)$ with the principal components. The results for electrons and photons injected at a broad range of energies are shown in Figs. \ref{fig:decaylimits} and \ref{fig:decaylimitband}.

If $\Delta  \Omega_\mathrm{dec} = \Omega_\mathrm{DM}$, i.e. the decaying species constitutes all the DM, we see that depending on the spectrum of decay products the lifetime is constrained to be $\tau \gtrsim 10^{23}-10^{25} s$, by \WMAP7 at the $95\%$ confidence level. For WIMP DM, this constraint is somewhat weaker than that obtained from present-day limits \cite{Dugger:2010ys}, from which we find $\tau \gtrsim 10^{26}$ s (albeit with substantial dependence on the decay channel). \emph{Planck} is expected to improve the limit from the CMB by nearly an order of magnitude. However, the general statement here, that if DM is still decaying in the present day then present-day probes are more sensitive than the CMB, should not be surprising. For $\tau \sim 10^{13}-10^{14}$s, the limit may be as strong as $\Delta \Omega_\mathrm{dec} \lesssim 10^{-11} \Omega_\mathrm{DM}$.

Comparing the forecast constraints for \emph{Planck} in the on-the-spot approximation, our results agree well with \cite{Zhang:2007zzh}, except for lifetimes smaller than $\sim 10^{13}$ s, where our limits are somewhat weaker. This may be due to a slightly different treatment of the corrections to recombination: in particular, our constraints are based on \cite{Finkbeiner:2011dx}, which ignored the production of additional Lyman-$\alpha$ photons due to excitations from the ionizing particles. This is conservative, in the sense that it weakens the constraints; the usual alternate treatment (see e.g. \cite{Chen:2003gz}) assumes that all energy going into excitations results in additional Lyman-$\alpha$ photons, which may overestimate the effect. The difference in the constraints is small for conventional DM annihilation (see the discussion in Appendix A of \cite{Finkbeiner:2011dx}), but this effect is most important around recombination itself: thus we might expect it to have the greatest impact in models where the decay lifetime is less than the age of the universe at recombination (since there are no significant constraints from decays when the universe was ionized).

\subsection{Comparison to previous bounds}

\begin{figure}
\includegraphics[width=0.45\textwidth]{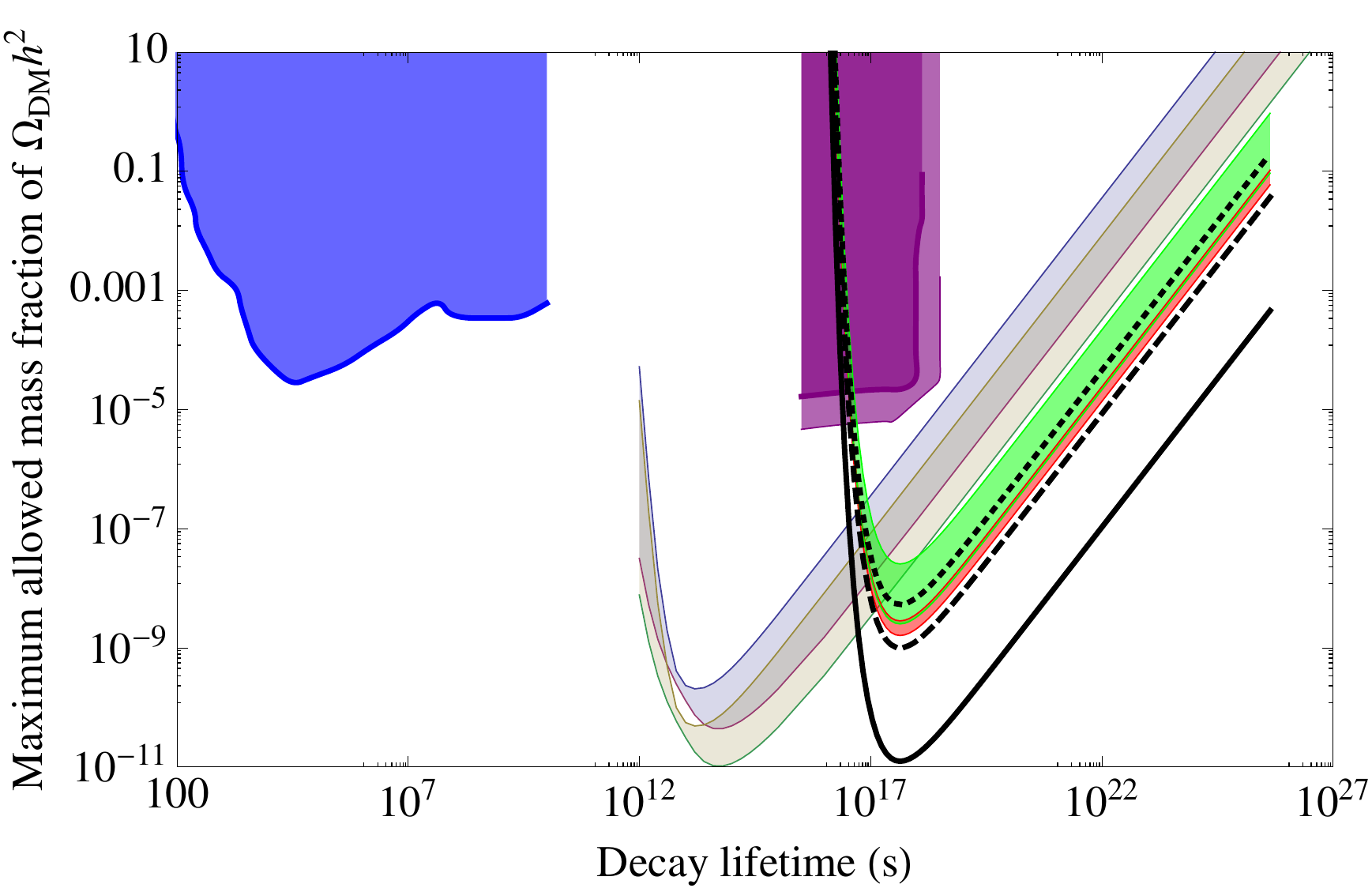}
\caption{\label{fig:otherdecaybounds}
As in Fig. \ref{fig:decaylimitband}, but also showing some other constraints from the literature on the parameter space. The blue region shows the sensitivity of BBN to a massive gravitino decaying hadronically into neutralinos \cite{Cyburt:2009pg}. The purple regions show the effect of a nearly-degenerate DM excited state on structure formation; the dark purple region is ruled out, whereas the light purple band has interesting deviations from CDM behavior \cite{Peter:2010au}. The light green band shows the sensitivity of the \emph{Fermi} Gamma-Ray Space Telescope to weak-scale DM decaying completely into $\mu^+ \mu^-$ pairs, using the combined constraints from clusters and the extragalactic background \cite{Dugger:2010ys}; all regions above the band are excluded, the width of the band corresponds to a scan over DM masses from 300 GeV to 10 TeV. The light red band shows the same result for complete decay into $b \bar{b}$ pairs. The black lines show the best possible current reach of experiments searching for the decay products directly, for decay into photons (\emph{solid}), electrons (\emph{dashed}) and neutrinos (\emph{dotted}), scanning the energy of the decay products from 4 keV to 60 TeV \cite{Bell:2010fk}.
}
\end{figure}

There are existing constraints on decaying metastable species from big bang nucleosynthesis, large-scale structure formation, and present-day indirect detection experiments. For comparison, we show some of these limits in Fig. \ref{fig:otherdecaybounds}. 

Peter et al \cite{Peter:2010au, Peter:2010jy, Peter:2010sz, Peter:2010xe} have studied the case where the DM has a nearly-degenerate excited state, which decays at some late time, providing a velocity ``kick'' to the DM particle. These kicks can modify the evolution of DM structure, and so such models can be constrained by comparison with observation. There is also a potentially interesting region where the resulting modifications to halos are not ruled out, but do not precisely reproduce the results of standard cold DM cosmologies. We show both the constrained and the ``potentially interesting'' region, and note that both appear to be ruled out by CMB constraints \emph{if} the decays have a significant branching ratio to electromagnetically interacting particles.

Injection of either leptons or hadrons can modify big bang nucleosynthesis, and constrain decaying species with lifetimes much smaller than the age of the universe at recombination. These constraints depend rather strongly on the decay channel; they are much more powerful for hadronic modes than leptonic modes. We do not present an exhaustive survey of such constraints, but show one illustrative bound, for a scenario where a massive gravitino decays hadronically into neutralino DM \cite{Cyburt:2009pg}. For decay lifetimes comparable to (and longer than) the age of the universe at recombination, rather than at BBN, the CMB constraints are far more sensitive.

If the lifetime of the decaying species is comparable to the \emph{present} age of the universe, a wide variety of probes from indirect detection experiments become viable; for example, studying gamma-rays in the isotropic diffuse photon background \cite{Yuksel:2007dr,Zhang:2009ut}, in the Milky Way \cite{Zhang:2009ut}, and in galaxy clusters \cite{Dugger:2010ys}. These limits depend rather strongly on the spectrum of decay products, and thus on the mass of the decaying species and whether all its mass energy is liberated in the decay. We show the limits on the maximum allowed mass fraction of the DM, as a function of lifetime, assuming a particle decaying completely to $b \bar{b}$ or $\mu^+ \mu^-$ pairs, and scanning over DM masses from 300 GeV to 10 TeV, using the combined constraints from clusters and the extragalactic background \cite{Dugger:2010ys}.

Going beyond gamma rays, the limits on decays into photons, electrons and neutrinos have been studied by \cite{Bell:2010fk}. The authors consider the case where the decay liberates only part of the particle's mass energy; as previously, the constraints are highly dependent on the spectrum. We plot the \emph{most} optimistic bound (i.e. choosing the mass splitting that leads to the greatest observable signal), scanning over splittings from 4 keV to 60 TeV. This approach has advantages over the CMB limits in that it is sensitive to decays to neutrinos, but only certain ranges of the mass splitting can be constrained (because of the sensitivity to the spectrum of decay products).

In general, we see that if the DM is still decaying in the present day, constraints from observation of the decay products can be much more stringent than the CMB bounds (and can even constrain neutrinos, which the CMB does not); however, this statement does depend on the spectrum and species of the decay products. For lifetimes smaller than $\sim 10^{17}$ s, the CMB constraints dominate, providing bounds on the maximal mass density liberated in decays that are comparable to the best present-day constraints. While the on-the-spot approximation may break down, the constraints vary by less than an order of magnitude for photon and electron energies ranging from sub-keV to 5 TeV. (Of course, there is a model dependence in the branching ratio to these electromagnetically-interacting final decay products -- the limits should be rescaled by this branching fraction -- but our analysis shows that the dependence on the \emph{spectrum} of the decay products is modest.)

\subsection{Limits on asymmetric dark matter}

In the ``asymmetric dark matter'' scenario, the DM possesses an asymmetry in the early universe similar to the baryonic sector, with the ``dark antimatter'' annihilating away completely and the residual DM yielding the observed $\Omega_\mathrm{DM}$. These models need not have any DM annihilation signature at late times, but one can be generated if the symmetric component is repopulated. This could occur by decay of a metastable state \cite{Falkowski:2011xh}, or if the DM acquires a small Majorana mass which induces oscillations between the two states \cite{Cai:2009ia,Cirelli:2011ac,Tulin:2012re}. In such scenarios, annihilation switches back on at some late time; if the critical timescale is before reionization, the CMB can effectively constrain such models.

The time dependence of the abundance of the symmetric component, and hence of the annihilation rate, can be quite non-trivial \cite{Cirelli:2011ac, Tulin:2012re}; in scenarios with oscillation, it depends on the interplay between annihilation, oscillation timescale and scattering of the DM on the baryonic matter (which damps the oscillations via decoherence). Using the results we have presented here, any arbitrary known energy injection history can be constrained, but we will not attempt to present a detailed survey of the sample annihilation histories in the literature.

\begin{figure*}
\includegraphics[width=.45\textwidth]{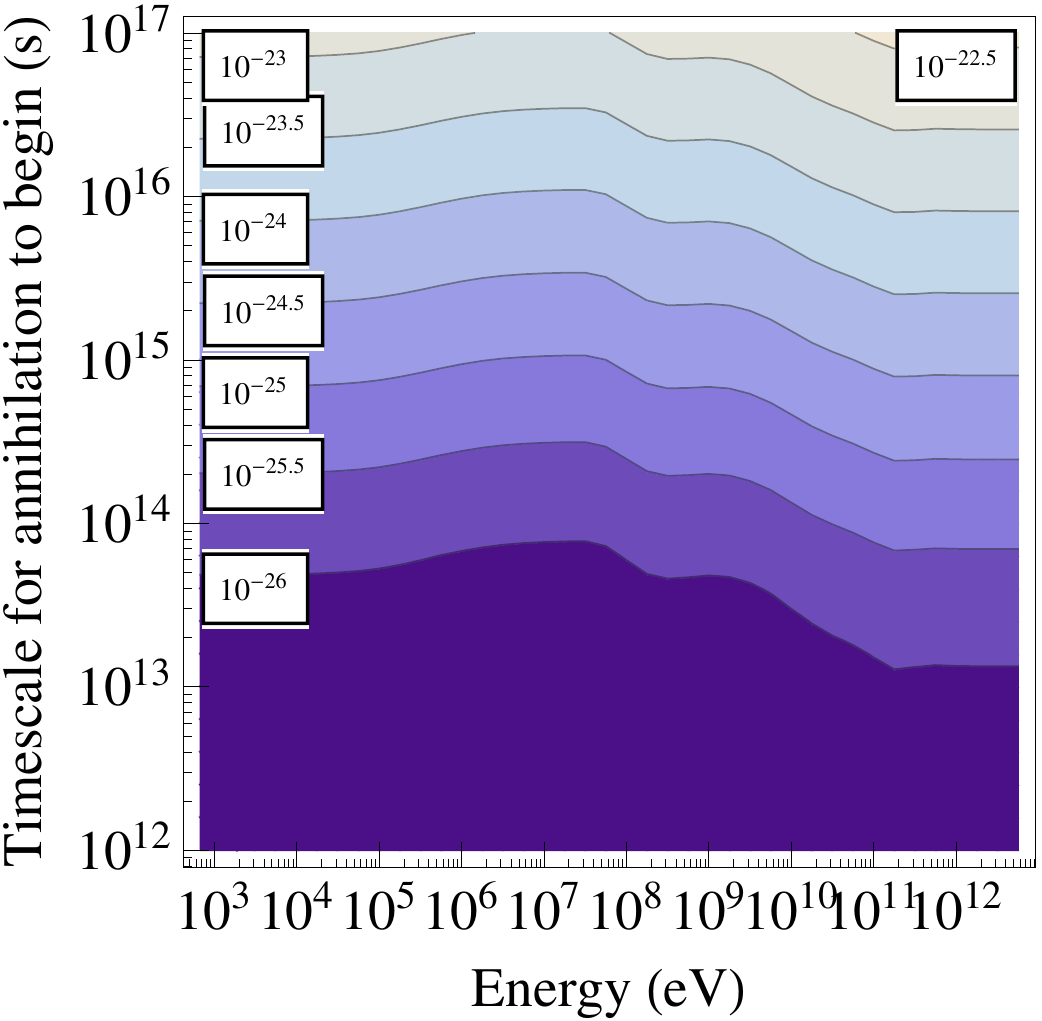}
\includegraphics[width=.45\textwidth]{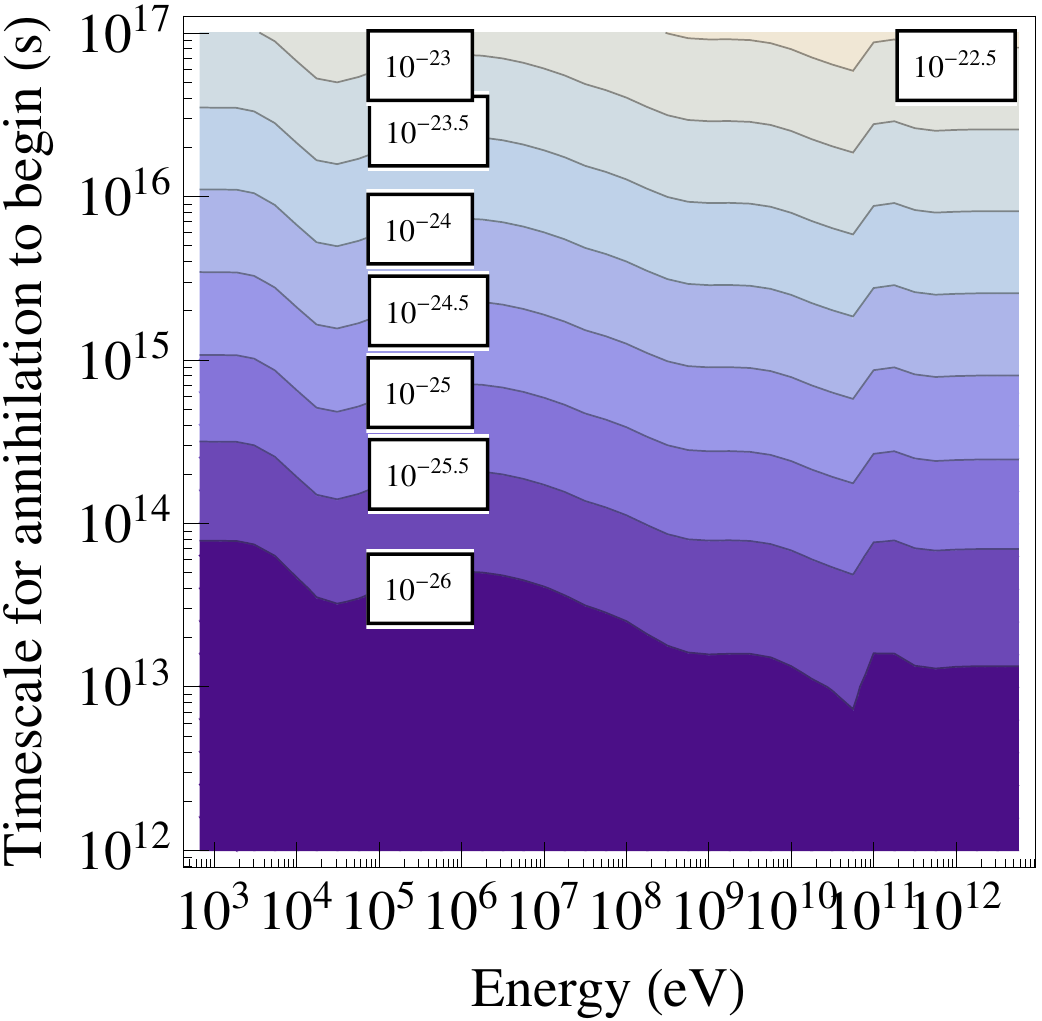} \\
\includegraphics[width=.45\textwidth]{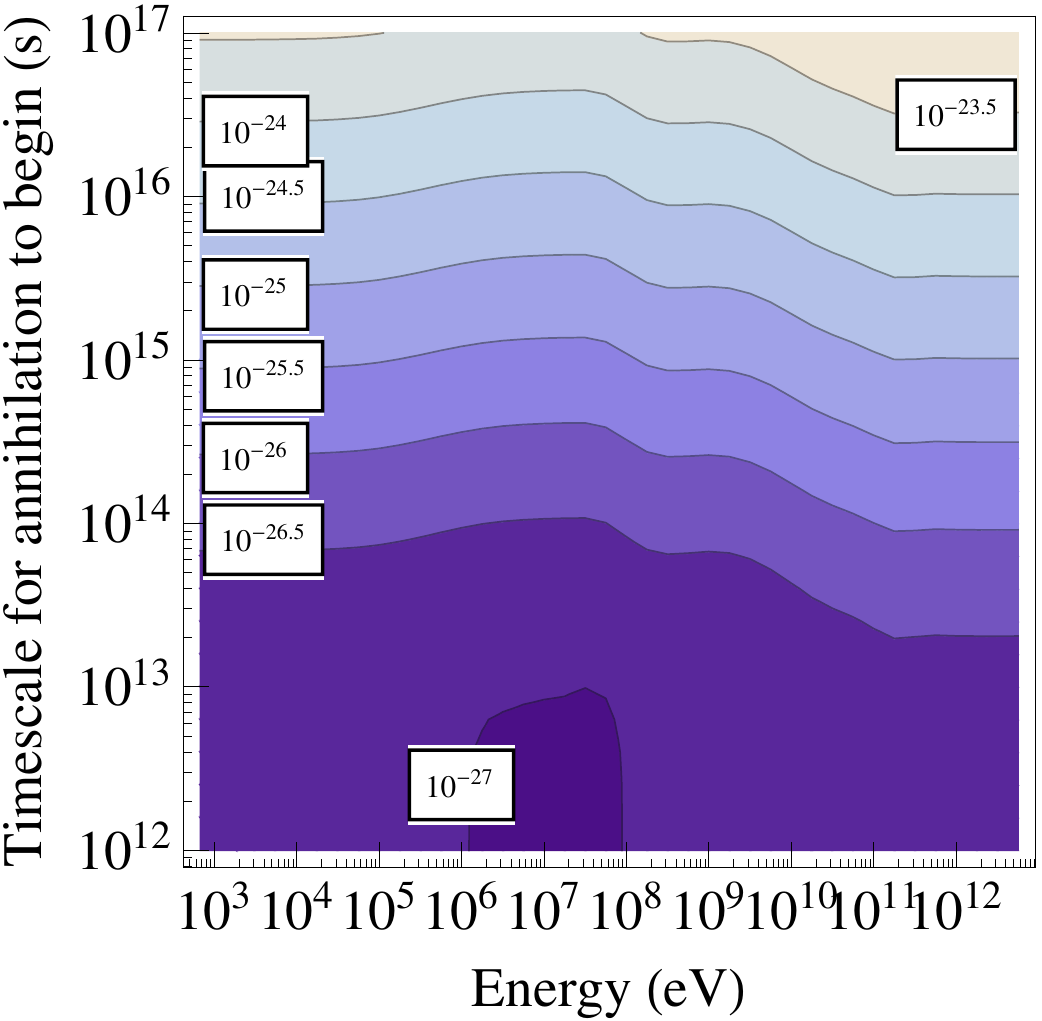}
\includegraphics[width=.45\textwidth]{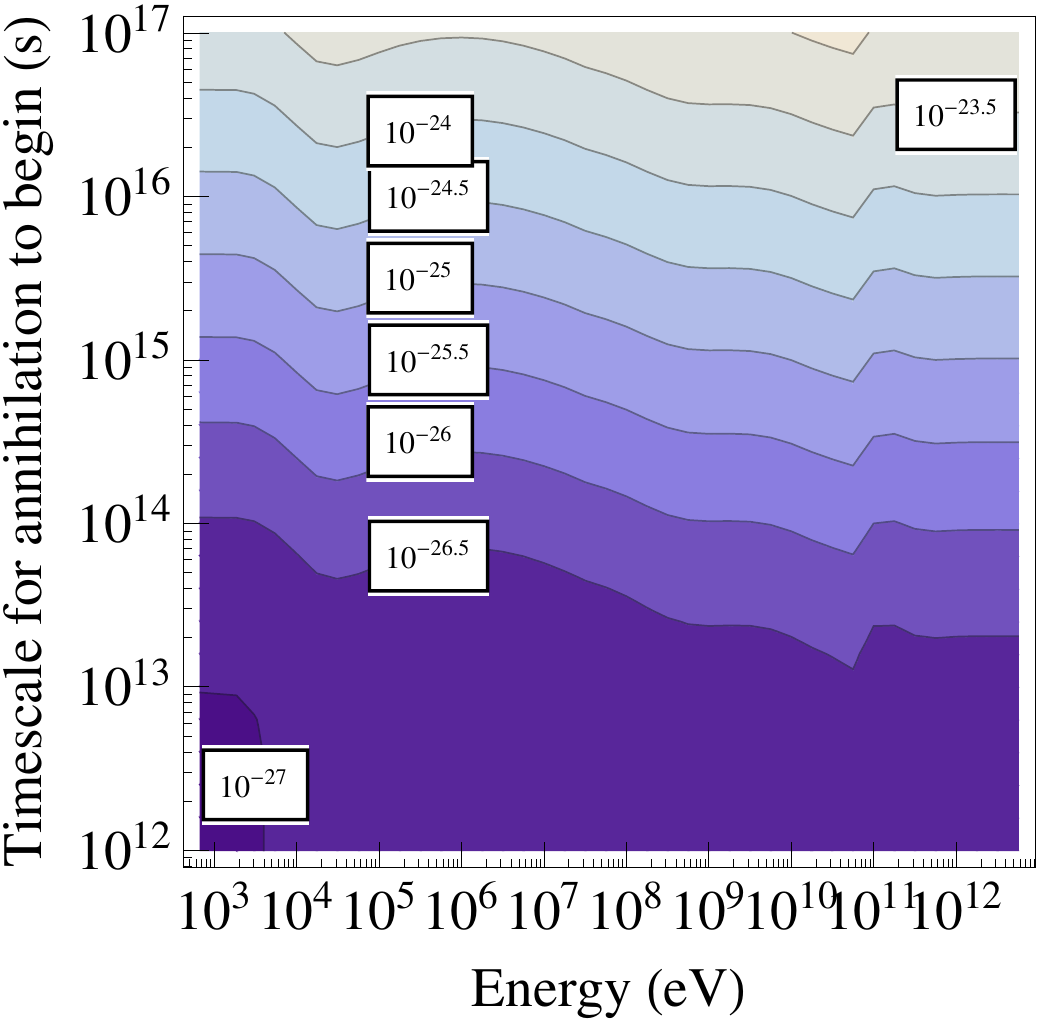}
\caption{\label{fig:osclimits}
Upper bounds on the parameter $\frac{\langle \sigma v \rangle}{m_\mathrm{DM}}$ for an asymmetric DM model where the annihilation turns on with some timescale $\tau$, as a function of $\tau$ and the energy of the decay products: the $95\%$ confidence limit from \WMAP7 (\emph{top row}) and the forecast $95\%$ confidence limit from \emph{Planck} (\emph{bottom row}). The left-hand panels apply to $e^+e^-$ pairs (as usual, the energy is for a single particle), the right-hand panels to photons.}
\end{figure*}

\begin{figure}
\includegraphics[width=0.45\textwidth]{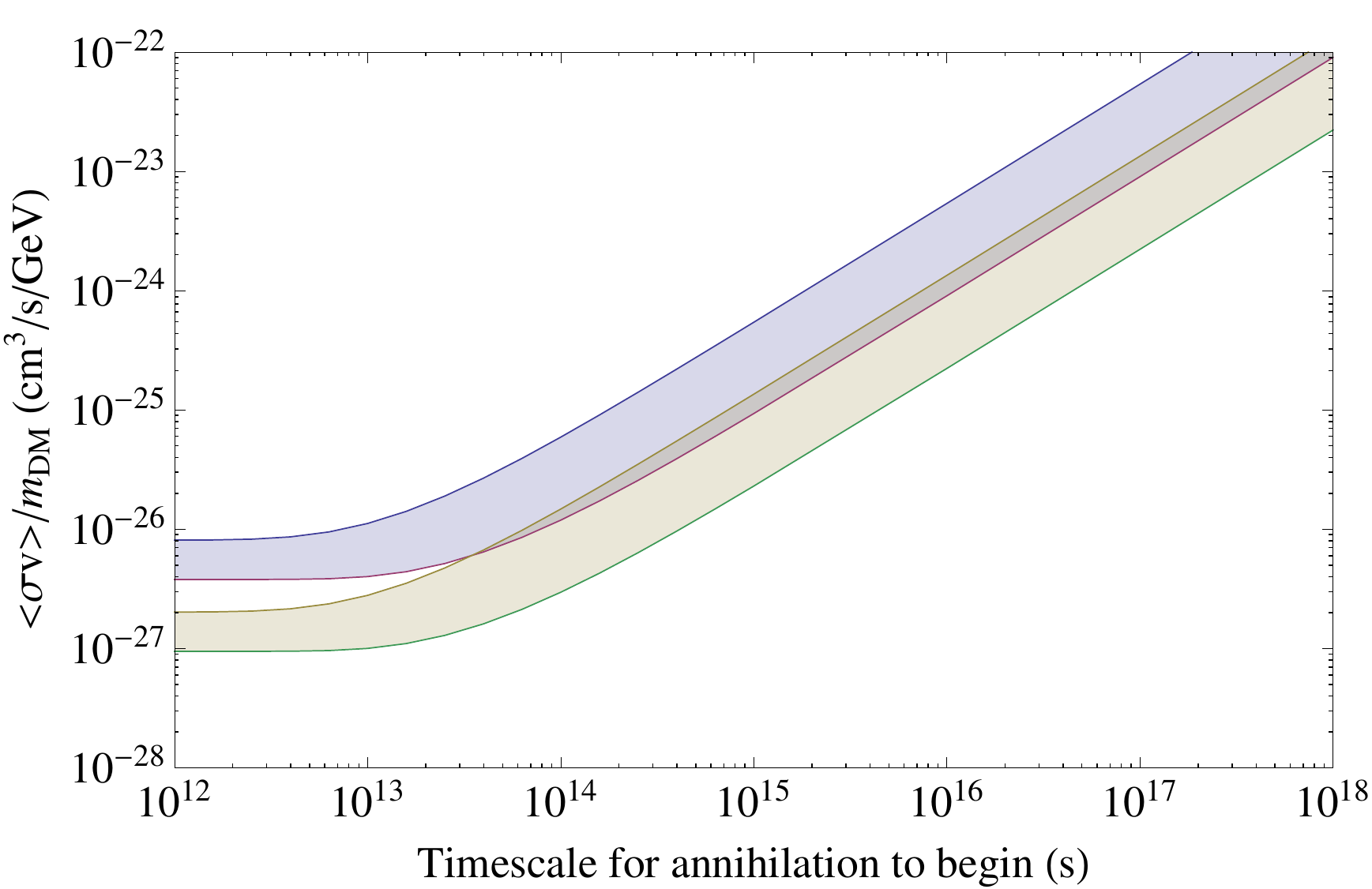}
\caption{\label{fig:osclimitband}
The $95\%$ confidence limits from \WMAP7 and \emph{Planck} (forecast) on the parameter $\frac{\langle \sigma v \rangle}{m_\mathrm{DM}}$ for an asymmetric DM model where the annihilation turns on with some timescale $\tau$, for a range of decay products and energies. For each timescale we scan over the constraint for photons and $e^+ e^-$ pairs with (single-particle) energy ranging from 1 keV to 6 TeV, to obtain the bands displayed; the width of the bands reflect the variation in the constraint between different energies and decay products. The upper (blue) band corresponds to \WMAP7 limits, the lower (yellow) band to \emph{Planck}. A branching fraction less than unity to electrons/positrons/photons will shift these bands upward. Values above the lines are excluded.
}
\end{figure}

Instead, as an example, we present bounds on a scenario where the annihilation rate scales as $1 - e^{-t/\tau}$ for some timescale $\tau$. This time dependence can occur in scenarios where the annihilation requires a component that is populated by the decay of another species, with timescale $\tau$; it is also applicable to ``oscillating asymmetric DM'' scenarios where DM scattering damps the oscillations, with fractional abundances for the ``DM'' and ``anti-DM'' states scaling approximately as $(1 + e^{-t/2 \tau})/2$ and $(1 - e^{-t/2 \tau})/2$, respectively. The resulting limits may also be approximately applicable to energy injection histories with large oscillations in the annihilation rate, since sufficiently rapid oscillations in the energy injection history should cancel out in their effect on the energy deposition history, the cosmic ionization history, and the CMB. However, this should be checked explicitly if precise limits are desired, once the energy injection history is known.

Since we are now dealing with DM particles that can only annihilate with their antiparticles, the annihilation rate per unit volume is given by $\langle \sigma v \rangle n_\mathrm{DM} n_\mathrm{\bar{DM}} \rightarrow \langle \sigma v \rangle n_\mathrm{tot}^2/4$ in the symmetric limit, where $n_\mathrm{tot} = n_\mathrm{DM} + n_\mathrm{\bar{DM}}$ is the total DM density. In comparison, for Majorana DM the corresponding rate is $ \langle \sigma v \rangle n_\mathrm{tot}^2/2$. Since the energy injected per annihilation is $2 m_\mathrm{DM}$, the energy injection rate per unit volume is given by,
\begin{equation} \left(\frac{dE}{dt dV} \right)_\mathrm{injected} = (1+z)^6 \Omega_\mathrm{DM}^2  \rho_c^2 c^2  \frac{ \langle \sigma v \rangle}{2 m_\mathrm{DM}} \left(1 -  e^{-t/\tau} \right), \end{equation}
and we define $f(z)$ by,
\begin{align} \left(\frac{dE}{dt dV} \right)_\mathrm{deposited} & = f(z) \varepsilon b(z), \nonumber \\
b(z) &  \equiv (1+z)^6 \Omega_\mathrm{DM}^2  \rho_c^2 c^2, \nonumber \\ \varepsilon & \equiv \frac{ \langle \sigma v \rangle}{2 m_\mathrm{DM}}. \end{align}
In the on-the-spot approximation, we would find $f(z) = 1 -  e^{-t(z)/\tau}$. In the limit where $\tau \ll t_\mathrm{recombination}$, i.e. the DM has oscillated or decayed to a purely symmetric phase by the onset of recombination, we should recover the results for ordinary annihilating Dirac DM.

Our limits on this scenario are shown in Figs. \ref{fig:osclimits}-\ref{fig:osclimitband}. For small $\tau$ the limit on $\frac{\langle \sigma v \rangle}{m_\mathrm{DM}}$ asymptotes to a fixed value, as expected; for the energy ranges where deposition is most efficient (corresponding to $f \approx 1$), the $95\%$ confidence bounds in the small-$\tau$ limit are,
\begin{align} \frac{\langle \sigma v \rangle}{m_\mathrm{DM}} & \lesssim 4 \times 10^{-27} \mathrm{cm}^3\mathrm{/s/GeV} \quad \text{(\emph{WMAP} 7)}, \nonumber \\ \frac{\langle \sigma v \rangle}{m_\mathrm{DM}} & \lesssim 9 \times 10^{-28} \mathrm{cm}^3\mathrm{/s/GeV} \quad \text{(\emph{Planck})}. \end{align}
These results are in good agreement with those found in \cite{Finkbeiner:2011dx} and \cite{Hutsi:2011vx} (for the \WMAP7 case, they are slightly stronger than the limits derived using \texttt{CosmoMC} in \cite{Finkbeiner:2011dx}, because the principal component analysis relies on a linearity assumption which breaks down at the $\sim 20\%$ level at the \WMAP7 $95\%$ confidence limit). Note there is a factor-of-2 difference in the bound because the constraints in \cite{Finkbeiner:2011dx} are for Majorana DM, whereas these are for Dirac DM; the same value of $\langle \sigma v \rangle$ leads to twice the injected power for Majorana DM, and consequently those constraints appear to be a factor of 2 stronger.

\section{Conclusions}
\label{sec:conclusions}

I have presented the results of a comprehensive calculation for the partition between free-streaming photons and energy absorbed by the photon-baryon plasma, for electrons, positrons and photons injected into the universe during the cosmic dark ages. The results apply to particles with sub-keV to multi-TeV energies, and can be used to immediately compute the energy-absorption history and hence the effect on the ionization history and the CMB for any arbitrary source of electromagnetically interacting particles, with any redshift and energy dependence. 

These results allow easy computation of constraints on oscillating asymmetric DM, decaying species, and models for DM structure formation. I have extended previous CMB bounds on decaying species by showing how the energy-absorption rate varies according to the initial energy of injection, and showed that for decays with lifetimes around $\sim 10^{13}-10^{14}$s, \emph{Planck} will probe models where the fraction of the DM mass density liberated in decays is as small as $10^{-11}$. For oscillating asymmetric DM, I have taken one sample scenario and shown how the constraints weaken as the characteristic timescale for repopulation for the symmetric component becomes longer than the age of the universe at recombination; for early oscillations, we recover the usual CMB bounds on DM annihilation. When DM structure formation is included, the fact that energy is not absorbed at the same redshift at which it is injected becomes very important for scenarios with annihilating DM; I have shown how the energy of the annihilation products changes the energy-absorption history (which will in turn modify the effect on reionization).

The complete set of numerical results used in this note, and a \texttt{Mathematica} notebook demonstrating their use, are available online (at \texttt{http://nebel.rc.fas.harvard.edu/epsilon}) for the benefit of the research community.

\section{Acknowledgements}
TRS is supported by NSF grants PHY-0969448 and AST-0807444.

\onecolumngrid

\begin{appendix}
\section{Online Supplemental Material}
\label{app:download}

We have made the tables $T_{e^+e^-}^{ijk}$ and  $T_{\gamma}^{ijk}$ (see Sec. \ref{sec:numerics}) available at \texttt{http://nebel.rc.fas.harvard.edu/epsilon}, in \texttt{.hdf} and \texttt{.fits} format, labeled as \texttt{resultsgrid\_elec.hdf} (\texttt{.fits}) and \texttt{resultsgrid\_phot.hdf} (\texttt{.fits}) respectively. Each \texttt{.hdf} or \texttt{.fits} file contains six arrays, labeled:

\begin{itemize}
\item \texttt{OUTPUT\_REDSHIFT}: this array provides the abscissa for output redshift, i.e. the value of $1+z$ at which the energy is deposited.
\item \texttt{LOG10(ENERGY/EV)}: this array provides the abscissa in energy, with values given by $\log_{10}(\text{energy in eV})$. Note that this is \emph{kinetic} energy; a particle annihilating or decaying to $e^+ e^-$ would need a mass sufficient to provide this energy in addition to the mass energy of the pair.
\item \texttt{INPUT\_REDSHIFT}: this array provides the abscissa for input redshift, i.e. the value of $1+z$ at which the energy is injected.
\item \texttt{DEPOSITION\_FRACTIONS}: this array provides the table $T^{ijk}$ for the appropriate species: that is, for a particle injected at some input redshift and energy (given by the abscissa arrays), the fraction of its initial energy deposited in the (log-spaced) timestep associated with the output redshift.
\item \texttt{F\_CHECK}: this array is redundant and intended only as a cross-check; it describes the effective efficiency $f$-curve for DM annihilation to the species in question, with DM mass given by \texttt{LOG10(ENERGY/EV)}, sampled at the redshift points given by the \texttt{OUTPUT\_REDSHIFT} array. It can be used to confirm that you are computing the effective f-curve correctly from the \texttt{DEPOSITION\_FRACTIONS} array.
\item \texttt{CONVERSION\_FACTOR}: this array does not use any of the results of the code, and is defined as $H(z)/(1+z)^3$, where $1+z$ is given by the  \texttt{INPUT\_REDSHIFT} array. It is provided for convenience only, as it is the weighting factor applied to the \texttt{DEPOSITION\_FRACTIONS} array to compute the $f(z)$ curves for DM annihilation (see Eq. \ref{eq:annf}).
\end{itemize}

For convenience, we also provide \texttt{.hdf} counterparts of the \texttt{.fits} files described in \cite{Finkbeiner:2011dx}, which are required to estimate the CMB constraints (see Sec. \ref{sec:constraints}), and a \texttt{Mathematica} notebook that reads the \texttt{.hdf} files and performs the calculations shown in this article.

\end{appendix}


\bibliography{deposition}

\end{document}